\documentclass{aa}  

\usepackage{graphicx}
\usepackage{txfonts}
\usepackage[colorlinks=true,linkcolor=blue,citecolor=blue]{hyperref}%
\usepackage{xcolor}
\usepackage{braket}
\usepackage{subcaption}
\usepackage{subcaption}
\usepackage{booktabs}
\usepackage{multirow}
\usepackage{tabularx}
\usepackage{systeme}
\usepackage{verbatim}
\usepackage{tabularray}
\usepackage{soul}

\NewColumnType{M}[1]{Q[m,c,#1]} 

\begin{document}

   \title{Heavily Obscured AGN detection: a Radio vs X-ray challenge}
    \titlerunning{Heavily Obscured AGN detection: a Radio vs X-ray challenge}
   \subtitle{}

   \author{G. Mazzolari \thanks{\email{giovanni.mazzolari@inaf.it}}
          \inst{1,2},
          R. Gilli\inst{2},
          M. Brusa\inst{1,2},
          M. Mignoli\inst{2},
          F. Vito \inst{2},
          I. Prandoni \inst{4},
          S. Marchesi \inst{1,2,3},
          M. Chiaberge \inst{8,9},
          G. Lanzuisi \inst{2},
          Q. D'Amato \inst{5},
          A. Comastri \inst{2},
          C. Vignali \inst{2},
          K. Iwasawa\inst{6,7},
          C. Norman \inst{8,9}.
          }
\authorrunning{G. Mazzolari et al.}
   \institute{ Dipartimento di Fisica e Astronomia, Universit`a di Bologna, Via Gobetti 93/2, I-40129 Bologna, Italy
         \and
             INAF – Osservatorio di Astrofisica e Scienza dello Spazio di Bologna, Via Gobetti 93/3, I-40129 Bologna, Italy    
         \and 
            Department of Physics and Astronomy, Clemson University,  Kinard Lab of Physics, Clemson, SC 29634, USA
        \and
            INAF – Istituto di Radioastronomia, Via Gobetti 101, I-40129 Bologna, Italy   
        \and
             INAF - Osservatorio Astrofisico di Arcetri, Largo Enrico Fermi 5, I-50125 Firenze, Italy
        \and
            Institut de Ciències del Cosmos (ICCUB), Universitat de Barcelona (IEEC-UB), Martí i Franquès, 1, 08028, Barcelona, Spain
        \and 
            ICREA, Pg. Luís Companys 23, 08010 Barcelona, Spain
        \and
            Space Telescope Science Institute, 3700 San Martin Drive, Baltimore, MD 21218, USA
        \and
            Department of Physics and Astronomy, Johns Hopkins University, Baltimore, MD 21218, USA
            }
   \date{}
 
  \abstract
   {In the supermassive black hole (SMBH)-galaxy coevolution scenario, heavily obscured Active Galactic Nuclei (AGN)  represent a fundamental phase of SMBH growth during which most of the BH mass is accreted and the scaling relations with the host galaxy are set. Obscured nuclei are thought to constitute a major fraction of the whole AGN population, but their statistics and evolution across cosmic time are still highly uncertain. Therefore, it is pivotal to identify new ways to detect this vast and hidden population of growing SMBH. A promising way to select heavily obscured AGN is through radio emission, which is largely unaffected by obscuration and can be used as a proxy for nuclear activity. }
   {In this work, we study the AGN radio detection effectiveness in the major deep extragalactic surveys, considering different AGN obscuration levels, redshift, and AGN bolometric luminosities. We particularly focus on comparing their radio and X-ray detectability, making predictions for present and future radio surveys.}
   {We extrapolate the predictions of AGN population synthesis model of cosmic X-ray background (CXB) to the radio band, by deriving the 1.4 GHz luminosity functions of unobscured (i.e. with hydrogen column densities $\log N_{H} <22$), obscured ($22<\log N_{H}<24$) and Compton-thick (CTK, $\log N_{H} >24$) AGN. We then use these functions to forecast the number of detectable AGN based on the area, flux limit, and completeness of a given radio survey and compare it with the AGN number resulting from X-ray predictions.}
   {When applied to deep extragalactic fields covered both by radio and X-ray observations, we show that, while X-ray selection is generally more effective in detecting unobscured AGN, the surface density of CTK AGN radio detected is on average $\sim 10$ times larger than the X-ray one, and even greater at high redshifts, considering the current surveys and facilities.\\ 
   Our results suggest that thousands of CTK AGN are already present in current radio catalogs, but most of them escaped any detection in the corresponding X-ray observations.\\
   We also present expectations for the number of AGN to be detected by the Square Kilometer Array Observatory (SKAO) in its future deep and wide radio continuum surveys, finding that it will be able to detect more than 2000 AGN at $z>6$ and some tens at $z>10$, more than half of which are expected to be CTK.}
   {}

   \keywords{galaxies: active – galaxies: evolution - radio continuum: galaxies – galaxies: luminosity function, mass function}

   \maketitle
%

\section{Introduction}\label{sec:intro}
Obscured Active Galactic Nuclei (AGN) are important astrophysical sources in the context of the supermassive black hole (SMBH)-galaxy coevolution scenario \citep{hickox18,Hopkins08}. In this framework, most SMBH and galaxy growth occurs during a very active obscured phase of black-hole accretion and star formation (SF). The radiative and kinetic energy released by the AGN is then supposed to sweep away the obscuring material, eventually quenching both SF and SMBH growth \citep{lapi18}.
A number of results from deep X-ray surveys indicate that the role of obscured and heavily obscured AGN might be particularly important at high redshift. \cite{vito18} showed that the fraction of AGN obscured by gas column densities ($\rm N_H$) larger than $\log N_H>23$ increases up to $\sim 80\%$ at z$\sim 4$, as also supported by analytical models and simulations \citep{ni20,lapi20,gilli22}. Furthermore, the SMBH accretion rate density (BHARD) derived by different coevolutionary models \citep{volonteri16, sijacki15,shankar14} is in good agreement with the observational results at least up to $z<3$, whereas at larger redshifts the models seem to overpredict the BHARD derived from X-ray surveys. While it is not clear yet whether issues in the simulations are responsible for this overprediction \citep{vito16}, one of the most widely proposed solutions is the existence of a highly obscured AGN population at high redshifts, missed by X-ray surveys \citep{barchiesi21}.\\
JWST is pushing the frontier of the spectroscopic identification of obscured AGN well into the epoch of the reionization, enabling us to look at the primordial stages of the SMBH and galaxy formation. Very recent results coming from JWST observations \citep{maiolino23,greene23} show a surprisingly high number density of broad-line (BL) AGN among the population of high-z compact red sources, which usually lack any X-ray detection even in deep X-ray fields \citep{ubler23}. In \cite{yang23} the identification of high-z obscured AGN using the reddest JWST filters led to a BHARD substantially higher than the X-ray results for $z>3$, suggesting that we are probably still missing the bulk of the AGN population at these redshifts. \\
Therefore it is of primary importance to detect and identify obscured Compton-thin  ($22<\log N_H<24$) and Compthon-thick (CTK; $\log N_H>24$) AGN, to understand their role in SMBH-galaxy coevolution and to constrain their physical properties and demographics, especially in the early Universe. \\
Among the several approaches reported in the literature to select AGN, the most commonly used are: X-ray selection \citep{ranalli05}, Near-infrared (NIR) and Mid-infrared (MIR) colour selection \citep{donley12}, Spectral Energy Distribution (SED) fitting \citep{boquien19,yang20,brammer08}, spectroscopic classification \citep{mignoli13}, and deviation from the Far-Infrared-Radio-Correlation of Star-Forming Galaxies (SFG) \citep{novak17,delvecchio17,delvecchio21}.\\
The first three diagnostics are normally used to select moderate-to-high luminosity AGN (HLAGN), i.e. radiatively efficient AGN usually characterized by bolometric luminosities larger than $\rm L_{bol}>10^{43} erg\ s^{-1}$. However, when heavily obscured AGN are taken into account, the effectiveness of X-ray and NIR-MIR selections shows severe limitations.\\
AGN obscuration is produced by dust and gas close to the central SMBH (the torus, in the inner parsec), but also by the cold interstellar medium (ISM) of the host galaxy \citep{aravena2020,decarli23,lusso23}. While optical emission is generally absorbed by dust, X-ray emission can penetrate through moderate-to-large gas column densities, but the X-ray detection of CTK AGN is challenging: for these objects, the X-ray emission in the energy range 0.5-10 keV is almost completely absorbed by gas, and only very high energy X-ray photons ($>10$keV) can escape. Moreover, X-ray selection suffers from the limited area-sensitivity combination of current X-ray facilities \citep{gilli22b}, making the detection of the faintest sources even harder.\\
Different NIR-MIR colour selection techniques proved to be effective in selecting obscured AGN by detecting the emission from their warm dusty torus \citep{stern12,assef13,donley12}. However, the fraction of contaminants (mainly dusty SFG) that mimic the colours of obscured AGN increases at high redshift, particularly for those selections relying on NIR photometric bands \citep{donley12}. Therefore, applying a cut in redshift ($z<3$) is usually necessary to provide reliable selections. Furthermore, NIR-MIR colour prescriptions need a strong AGN component; hence, systems with weak AGN emission are not easily identified \citep{hickox18}.\\
Radio emission can uncover low-luminosity AGN, that do not show the strong signatures of HLAGN, and with less time-demanding observations compared to spectroscopy or X-ray. A great advantage of radio emission is that it is largely unaffected by AGN obscuration since both gas and dust opacities are almost negligible at typical radio frequencies \citep[e.g. 1.4 GHz, ][]{hildebrand83}.\\
Historical radio surveys were mostly sensitive to the powerful ($\rm L_{1.4\, GHz}>10^{25} \ W \, Hz^{-1}$)  population of the so-called radio-loud (RL) AGN \citep{white97,becker95,becker01}, for which a significant fraction of the power produced by the accretion process is released in kinetic form through the formation of relativistic jets that may expand up to the Mpc scales. However, new-generation radio surveys \citep{heywood20,alberts20,vandervlugt20,hale23} are deep enough to detect also SFG and so-called radio-quiet (RQ) AGN, i.e. radiatively efficient AGN with much weaker radio emission, typically confined on (sub-)kpc scales. In particular, reaching $\mu \rm Jy$ sensitivities, it is possible to detect radio luminosities down to $\rm L_{1.4\, GHz}\sim 10^{23} \, \rm W \, Hz^{-1}$ (corresponding to $\nu L_{\nu}\sim 10^{39} \, \rm erg \, s^{-1}$; with $\rm \nu=1.4GHz$) and star-formation rates SFR $\sim 100 \, \rm M_{\odot} \, yr^{-1}$ up to $z\sim 3$.\\ 
At frequencies around a few GHz, the radio emission is, both for AGN and SFG, mostly due to optically thin synchrotron radiation, emitted by electrons accelerated to relativistic velocities by different acceleration mechanisms. In SFG the acceleration is provided by supernova explosions in SF regions, while in RL AGN the expanding radio jets originate from the interaction of the electrons with the strong magnetic field amplified by the SMBH accretion disc \citep{blandford77}. In RQ AGN multiple mechanisms may be at work. Radio emission can be associated with circumnuclear SF regions, small (sub-)kpc scale jets, and/or other AGN-related mechanisms, like e.g. shocked radiative winds or coronal winds \citep[see ][for a comprehensive review]{panessa19}. A way to distinguish between SF or AGN-related radio emission is provided by the well-known Far Infrared Radio Correlation \citep{novak17,delvecchio21}. This relation arises because the same population of massive stars that heats up dust, causing it to reradiate its energy in the far-infrared (FIR), produces supernovae that generate relativistic particles emitting synchrotron radiation at radio frequencies. This correlation is quantified by a parameter $q_{\rm{TIR}}=\log(L_{TIR})/\log(L_{1.4GHz})$, i.e. the ratio between the total IR luminosity integrated between 8-1000$\mu$m and the 1.4GHz luminosity. This correlation has been found to depend both on the redshift and the stellar mass ($\rm M_{*}$) of the sources \citep{delhaize17, delvecchio17, delvecchio21}. Galaxies with radio emission in excess of what is predicted by the FIR-radio correlation likely possess an AGN-driven radio component, independently of which exact mechanism is responsible for it. 
It is common to indicate these sources as Radio-Excess (REX) AGN\footnote{REX AGN are a distinct class with respect to RL AGN, as in the former the observed radio emission is compared to the infrared emission due to star formation, while in the latter the radio emission is compared to the AGN-related emission observed at e.g. optical or X-ray bands (see  \cite{Kellermann1989,terashima03}). Nevertheless, RL AGN are also REX AGN as they also satisfy the REX definition.} \citep{delvecchio17,nandra07}.\\
In this work we aim to investigate the effectiveness of selection methods based on radio and X-ray emission considering AGN at different levels of obscuration and at different intrinsic luminosities and redshift, focusing particularly on CTK AGN. In Sec.~\ref{sec:methods} we translate the AGN X-ray luminosity functions into the radio band (1.4GHz) and predict the number of detectable AGN over a radio field, given its depth, area, and completeness corrections. In Sec.~\ref{sec:results} we present the 1.4GHz luminosity function (Sec.~\ref{sec:RLF}), radio number counts (Sec.~\ref{sec:NC_comp}) and AGN predictions (Sec.~\ref{sec:RX_field}), comparing the results with those in the literature. In Sec.~\ref{sec:Discussion} we apply our model to the major extragalactic fields covered by X-ray and radio observations, discussing for which type of AGN the 1.4GHz emission performs better than the X-ray one. In Sec.~\ref{sec:hz}, we focus on the predictions for the high-redshift Universe ($z>3$), while in Sec.~\ref{sec:SKA_pred} we present the expectations for the Square Kilometer Array Observatory (SKAO) surveys that will be done in the next future.\\

We assume a flat $\Lambda$CDM universe with $H_{0}=70 \rm{km s^{-1} Mpc^{-1}}$, $\Omega_{m}=0.3, \Omega_{\Lambda}=0.7, \Omega_{k}=0$.\\
We assume AGN radio spectra of the form $S_{\nu}\propto \nu^{-\alpha}$, with $\alpha=0.7$, which is the typical spectral slope considered for extragalactic synchrotron emission \citep{smolcic17a,novak17}. When $\rm L_{1.4\rm GHz}$ is reported we refer to $\rm \nu L_{\nu}$ (with $\rm \nu=1.4GHz$), in units of $\rm erg \, s^{-1}$. 

\section{Methods}\label{sec:methods}
Cosmic X-ray Background (CXB) models \citep{gilli07, ueda14, buchner15, ananna19}, by studying the integrated X-ray emission of faint extragalactic point-like sources, can provide an AGN census for any obscuration level and to include also the most CTK AGN population, poorly sampled even in the deepest X-ray surveys. The different CXB models almost agree in predicting a consistent fraction of CTK AGN among the whole AGN population, i.e. between $30-50\%$.\\    
To derive the AGN radio luminosity function (RLF) we followed the approach described in the following sections which is based on the CXB model of \cite{gilli07} (G07 hereafter) and considers the AGN radio-hard-X luminosity relation ($\rm L_{1.4\rm GHz}-L_{HX}$) derived by \cite{damato22}.\\
A similar approach was also followed in \cite{lafranca10}, to investigate the global AGN kinetic energy release in the context of AGN feedback processes.

\begin{figure}
	\includegraphics[width=1.0\columnwidth]{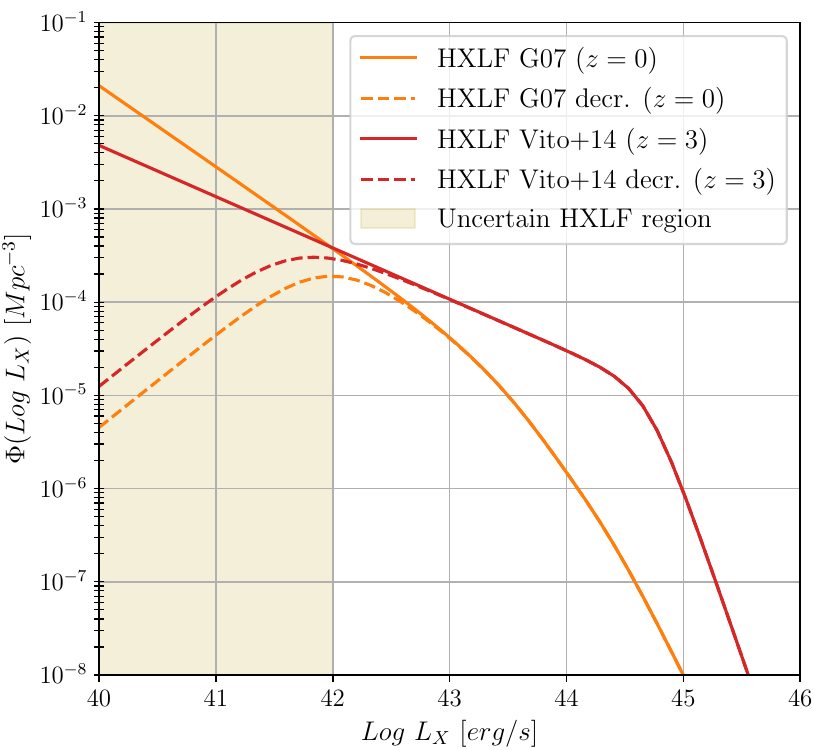}
    \caption{HXLF of all AGN obtained following the prescription of the CXB model in G07 and computed at $z=0$ (orange solid line). The orange dashed line is the same HXLF but introducing a cut-off at $\log L_X<42$, where the HXLF measurements are most uncertain. The red curve represents the HXLF as derived by \cite{vito14}, computed at $z=3$, and implemented as baseline model for $z>3$. The red dashed line is the \cite{vito14} HXLF with the same cut-off described above. }
    \label{fig:HX_LF}
\end{figure}

\subsection{Hard-X Luminosity function}\label{sec:HXLF}
The CXB model in G07 considers for the unobscured AGN population the soft-X-ray luminosity function (SXLF, energy range: 0.5-2 keV) derived by \cite{hasinger05}, of the form:
\begin{equation}\label{eq:phi_has_ext}
    \frac{d\Phi(L_X,z)}{d L_X}=A\biggl[\biggl(\frac{L_X}{L_*}\biggr)^{\gamma_1}+\biggl(\frac{L_X}{L_*}\biggr)^{\gamma_2}\biggr]^{-1} \cdot e_z(z,L_X) \cdot e_{\rm decl}(z).
\end{equation}
where the values of the normalization $A$, of the characteristic luminosity $L_{*}$, and of $\gamma_1,\gamma_2$ are taken from Tab. 5 of \cite{hasinger05}. The evolution factor $e_z$ was derived in \cite{hasinger05} considering a luminosity-dependent evolution model. The term $e_{\rm decl}$: 
\begin{equation}
    e_{\rm decl}(z)= 
    \begin{cases}
    1 \quad \text{for} \quad z<2.7 \\
    10^{[-0.43\cdot (z-2.7)]}\quad \text{for} \quad z\geq2.7
    \end{cases}
    \label{eq:z_decl}
\end{equation}
is introduced to reproduce the steep decline in the density of AGN observed at high-z \citep[e.g. ][]{Brusa2009}.\\
Following the same approach as in G07, we obtain the corresponding Hard-X luminosity function (HXLF, energy range: 2-10 keV) assuming a Gaussian distribution of X-ray spectral indices centred at $\Gamma=1.9$ \citep{piconcelli05}. Then, using the obscured-to-unobscured AGN ratio in G07, we derive the HXLF of the different sub-populations of AGN in terms of their level of obscuration. In particular, we derive the HXLF of obscured Compton-thin 
and CTK AGN, the latter assumed to have the same number density of the obscured Compton-thin AGN.
The orange solid line in Fig.~\ref{fig:HX_LF} corresponds to the total HXLF at $z=0$, computed by summing the luminosity functions of all the sub-populations of obscured and unobscured AGN.\\
The faint end of the XLF is poorly observationally constrained for values of $\log L_X\leq 42$, since at these X-ray luminosities it is difficult to distinguish between the X-ray emission coming from AGN and that coming from SF processes in normal galaxies. 
To derive information in the faintest AGN luminosity regime ($\log L_X\leq 42$), X-ray luminosity functions are usually extrapolated using the steep slopes determined at brighter luminosities, but the possibility that they remain constant or even decline cannot be excluded. Consequently, we introduced a cut-off in the \cite{hasinger05} SXLF at $\log L_X<42$:
\begin{equation}\label{eq:phi_has_decl}
    \frac{d\Phi(L_X,z)}{d L_X}=A\biggl[\biggl(\frac{L_X}{L_*}\biggr)^{\gamma_1}+\biggl(\frac{L_X}{L_*}\biggr)^{\gamma_2}+\biggl(\frac{L_X}{L_1}\biggr)^{\gamma_3}\biggr]^{-1} \cdot e_z(z,L_X) \cdot e_{\rm decl}(z).
\end{equation}
where $\rm L_1=10^{40}\ erg\ s^{-1}$ and $\gamma_{3}=-1$ (see orange dashed line in Fig.~\ref{fig:HX_LF}). \\
Furthermore, since the X-ray luminosity function of \cite{hasinger05} was largely derived from AGN samples at $z<3$, we also considered the high-z X-ray luminosity function presented in \cite{vito14}. In this work the authors, assembling a sample of 141 AGN at $3<z<5$ from X-ray surveys of different sizes and depths, built an HXLF specific for high-z AGN. In our work, we used the \cite{vito14} results as baseline HXLF at $z>3$. In \cite{vito14} HXLF we assumed a constant obscured AGN fraction with luminosity and a number ratios between unobscured, obscured Compton-thin, and
obscured Compton-thick AGNs of 1:4:4. Both these assumptions are in agreement with the observational results reported in \cite{vito18}. The total HXLF is presented in red in Fig.~\ref{fig:HX_LF} at $z=3$ (red solid line) and has the following analytic expression:
\begin{equation}\label{eq:phi_vito}
    \frac{d\Phi(L_X,z)}{d L_X}=A\biggl[\biggl(\frac{L_X}{L_s}\biggr)^{\gamma_a}+\biggl(\frac{L_X}{L_s}\biggr)^{\gamma_b}\biggr]^{-1} \cdot \Bar{e}_z(z,L_X)
\end{equation}
where the values of the parameters $L_s$, $\gamma_a$, $\gamma_b$, are taken from Table 5 in \cite{vito14} and the redshift evolution factor, $\Bar{e}_z(z,L_X)$ is given by Eq. 7 of the same work. The HXLF derived in \cite{vito14} provides a very good description of the high-z obscured AGN fraction and AGN space density measured in the 2Ms CDFN and 7Ms CDFS \citep[see ][]{vito18}.\\
In conclusion, our baseline model uses the HXLF derived by the CXB model in G07 for $z<3$ and of the HXLF of \cite{vito14} one for $z>3$. To account for the uncertainty on the X-ray luminosity function faint end also at high-z, we applied to the \cite{vito14} HXLF the same cut-off ($(L_X/L_1)^{\gamma3}$) reported in Eq.~\ref{eq:phi_has_decl}, (see red dashed line in Fig.~\ref{fig:HX_LF}).

\subsection{Predicted AGN Radio Luminosity Function}\label{sec:RLF}
To convert the HXLF into a RLF requires a relation between AGN luminosities in the two bands \citep[e.g.][]{panessa15,merloni03,dong21}. In this work, we assume the relation of \cite{damato22}, that was derived using the deep radio and X-ray data available in the $\rm \sim 0.2 deg^2$ field centred on the quasar SDSS J1030+0524 (hereafter J1030):
\begin{equation}\label{eq:LR_LX}
    \log L_{1.4\rm GHz}=0.83 \log L_{HX}+3.17.
\end{equation}
This relation was computed from a sample of X-ray selected AGN and early-type galaxies (ETG), spectroscopically confirmed up to $z\sim3$, spanning a wide range of luminosities. The 2-10 keV luminosities 
cover  the range $\rm \log L_{HX}\in [40.5,45]$, while the radio luminosities extend over $\log L_{1.4\rm GHz} \in [37,43]$.  Radio upper limits were taken into account using survival analysis. Since the relation was computed using X-ray selected sources, and since we started from an X-ray luminosity function, this relation is perfectly suited for our aim. Furthermore, since the sample of \cite{damato22} includes sources up to $z\sim 3$, their relation extends well beyond the local universe, as opposed to other $\rm L_{1.4\rm GHz}-L_{HX}$ relations in the literature.\\
The relation of \cite{damato22} does not consider any prior selection in the radio loudness parameter of the sources, but 83\% of the sample (87\% accounting radio upper limits) is radio quiet according to the radio loudness threshold $R_X=\log(L_{1.4\rm GHz}/L_{HX})<-3.5$, originally defined in \cite{terashima03}. The relation derived from \cite{damato22} differs significantly from those found in the literature for RL AGN \citep[e.g.][]{fan16}. This means that the RL AGN population 
cannot be described by this relation and they will not be included in our predictions (see Sec.~\ref{sec:results}).\\
To transform the HXLF into a RLF we considered a Gaussian probability distribution function that returns the probability $P$ that an AGN with a 2-10 keV luminosity $\rm L_{HX}$ has a radio luminosity $\rm L_{1.4GHz}$ according to the \cite{damato22} $\rm L_{1.4\rm GHz}-L_{HX}$ relation:
\begin{align} \label{eq:P_RLF}
    \rm P(L_{1.4GHz}&|L_{HX},\sigma_{\rm R}) = \\
    &=\frac{\exp(-(0.83 \log L_{HX}+3.17- \log L_{1.4GHz})^2 / 2\sigma_{\rm R}^2)}{\sqrt{2\pi}\sigma_{\rm R}},
\end{align}
where the Gaussian dispersion $\sigma_{\rm R}=0.5$ is given by the intrinsic dispersion of the $\rm L_{1.4\rm GHz}-L_{HX}$ relation. Then, we computed the RLF by weighing the HXLF by the probability distribution computed above: 
\begin{equation} \label{eq:RLF}
    \rm \Phi(L_{1.4GHz},z)=\int^{L_{HX_{max}}}_{L_{HX_{min}}} \Phi(L_{HX},z) \cdot P(L_{1.4GHz}|L_{HX},\sigma_{\rm R})\  dL_{HX}, 
\end{equation}
where $L_{HX_{max}}=10^{47}\rm erg\ s^{-1}$ and $L_{HX_{min}}=10^{40}\rm erg\ s^{-1} $. The AGN RLF computed in Eq.~\ref{eq:RLF} is shown in Fig.~\ref{fig:RLF_smolc} for  different redshift bins.\\
The shape of the RLF mainly depends on two parameters: the shape of the HXLF, in particular at its faint end, and the value of the dispersion $\sigma_{\rm R}$ of the $L_{1.4\rm GHz}-L_{HX}$ relation.\\ 
Our baseline model adopts the HXLF extrapolated to low luminosities using its original slope, combined with $\sigma_{\rm R}=0.5$ (i.e. the $1\sigma$ scatter of the X-ray-radio relation derived by \cite{damato22}). The shaded area in Fig.~\ref{fig:RLF_smolc} represents the assumed uncertainties on the RLF. The lower limit of the shaded region corresponds to the RLF computed starting from the HXLF with the cut-off at $\rm \log L_X<42$ (Eq.~\ref{eq:phi_has_decl}), and taking a slightly smaller dispersion $\sigma_{\rm R}=0.4$. The upper limit of the green shaded region is derived from the baseline model, but assuming a slightly larger dispersion $\sigma_{\rm R}=0.65$. \\
We note that a larger (lower) value of $\sigma_R$ with respect to our baseline model increases (decreases) the slope of the RLF, in particular at high $\rm L_{1.4GHz}$. 
The two values of the dispersion $\sigma_{\rm R}$ for the upper and lower boundaries are chosen to provide conservative estimates of the RLFs. The value $\sigma_{\rm R}=0.65$ is the one obtained from the \cite{damato22} sample when excluding radio upper limits;  $\sigma_{\rm R}=0.4$ allows us to obtain a steeper slope of the RLF at high radio luminosities.
\subsection{Radio Number Counts}
Starting from the RLF we computed the cumulative number counts (NC), namely the number of AGN above a given 1.4GHz flux $S_{lim}$ and in a given redshift range in units of $\rm sr^{-1}$ or $\rm deg^{-2}$. 
Taking the comoving volume element:
\begin{align}
   \frac{dV}{dz}=\frac{4\pi c}{H_{\rm 0}} \frac{D^{2}_{\rm M}}{\bigl(\Omega_{\rm m}(1+z)^3 + \Omega_k (1+z)^2 + \Omega_{\rm \Lambda}\bigr)^{1/2}},
\end{align}
where $D_{\rm M}$ is the comoving distance, the number of AGN per steradian with a flux $\rm S>S_{lim}$ is given by:
\begin{align}\label{eq:NC}
    N(>S_{lim})=&\int \frac{dN}{ds}ds\\
    &=\frac{1}{4\pi} \int^{z_{max}}_0\int^{L_{max}}_{max[l_{lim},L_{min}]}\Phi(L,z)\cdot \frac{dV}{dz} \ dL \ dz 
\end{align}
where $l_{lim}=L_{1.4GHz}(S_{lim},z)$, and we fixed $z_{\rm max}=10$, $L_{\rm max}=10^{43}\rm erg\ s^{-1}$ and $L_{\rm min}=10^{37}\rm erg\ s^{-1}$. 

\section{Results}\label{sec:results}

\subsection{Luminosity function comparison}\label{sec:LF_comp}

\begin{figure*}
	\includegraphics[width=2\columnwidth]{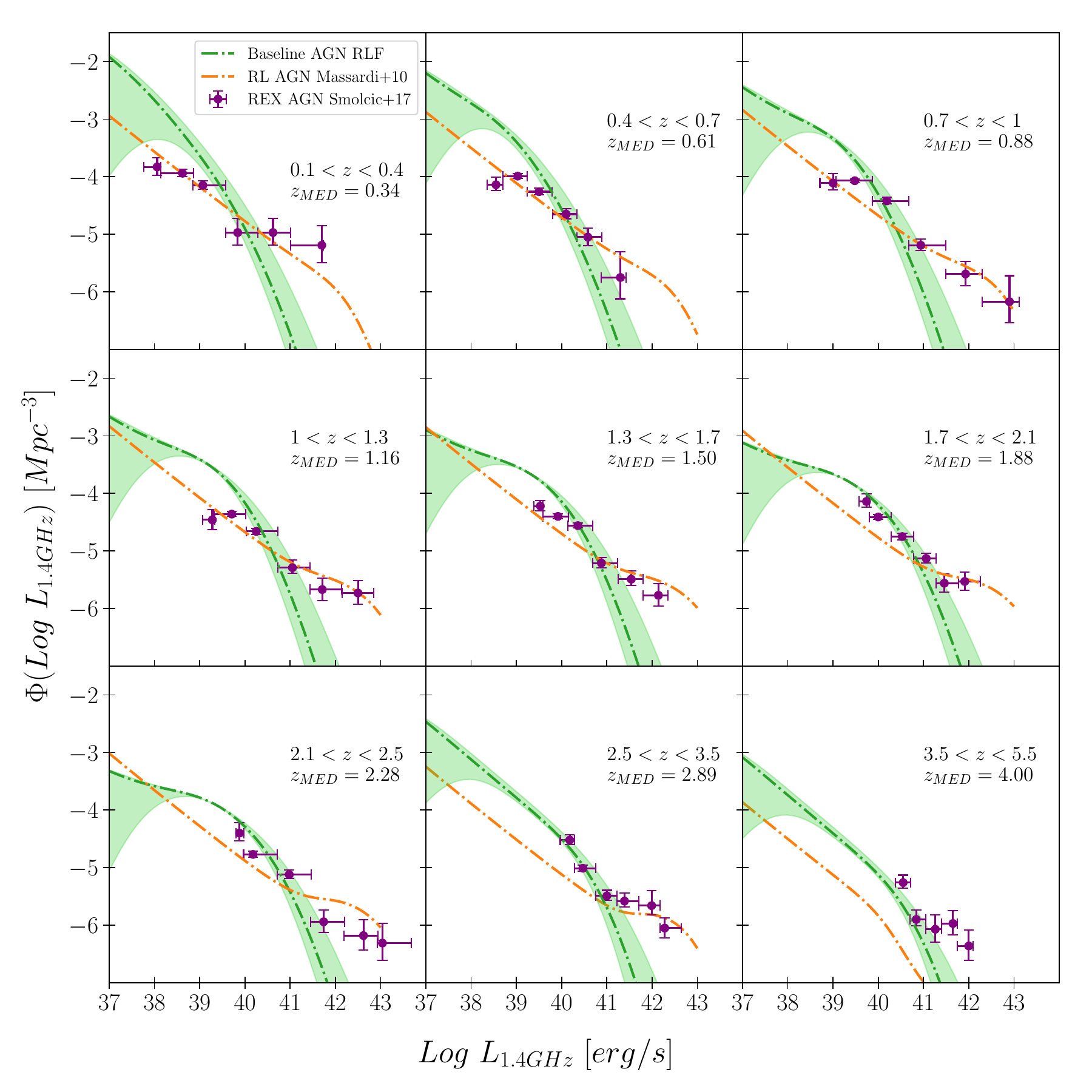}
    \caption{Baseline RLF of all AGN derived from our model (green line) compared to the REX AGN RLF (purple data points) measured by \citet{Smoilcic17c}. Different redshift ranges are investigated, as labelled. The boundaries of the shaded area represent the uncertainty region of the RLF as described in Sec.~\ref{sec:RLF}. In all panels we also report in orange the RL luminosity function empirically derived by \cite{massardi10}.}
    \label{fig:RLF_smolc}
\end{figure*}

\begin{figure*}
\centering
	\includegraphics[width=2\columnwidth]{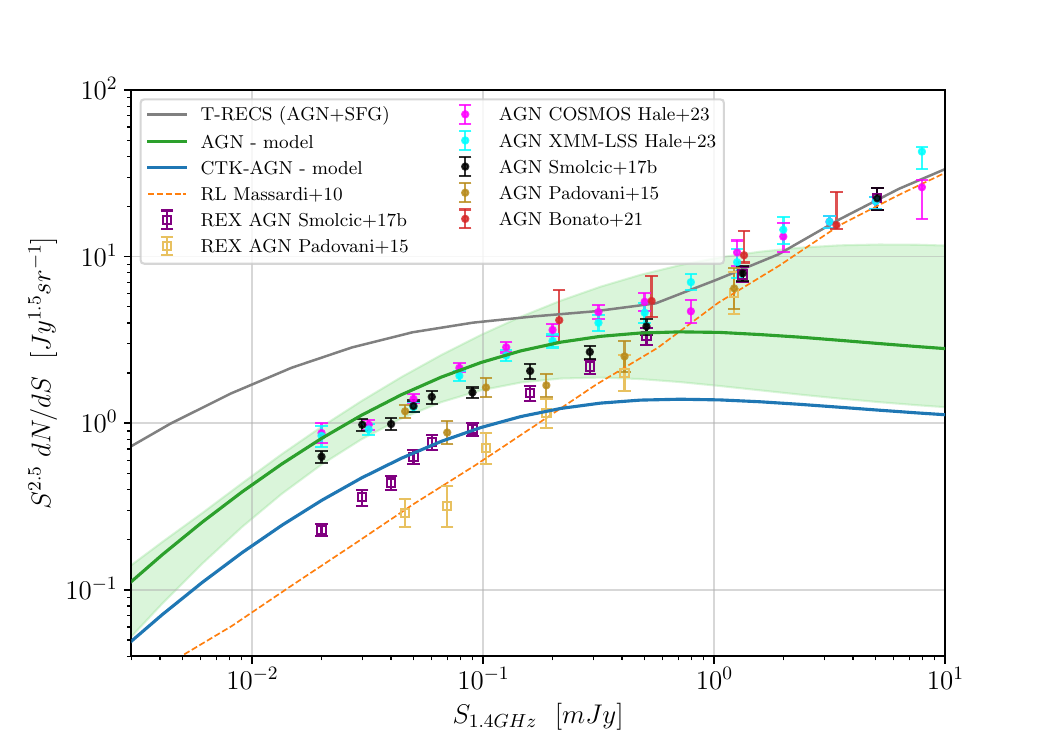}
    \caption{Differential AGN number counts (NCs) derived with our model for all AGN (green line and shaded area) and CTK AGN (blue curve), compared with AGN NCs derived in the literature from different radio fields, as labelled. The uncertainties on the CTK NCs are of the same order as those of the whole AGN population but are not reported for clarity. NCs measured for the whole radio AGN population in the different works are shown as filled points, while the REX AGN only are shown as empty squares. The grey curve shows the NCs of AGN plus SFG predicted by the T-RECS simulations, while the orange curve shows the radio-loud AGN NCs derived from \cite{massardi10}.} 
    \label{fig:number_counts}
\end{figure*}

In Fig.~\ref{fig:RLF_smolc} we present the RLF obtained with our model compared to that measured in the COSMOS field for REX AGN at different redshifts \citep{Smoilcic17c}.\\
The COSMOS REX AGN are selected as radio sources exhibiting a >$3\sigma$ radio emission excess with respect to the one expected from their hosts (IR-based) SFR  \citep{Smoilcic17c}. This criterion ensures that at least 80\% of the radio emission in these sources is associated with an AGN, resulting in a highly reliable sample.\\
Fig.~\ref{fig:RLF_smolc} shows that our model generally predicts a higher AGN number density at low radio luminosities with respect to \cite{Smoilcic17c}, while it predicts lower AGN number densities at high radio luminosities. On the other hand, the modelled RLFs are consistent with the data (within the error bars) in the intermediate luminosity range. The fact that our model cannot reproduce the high-radio luminosity AGN number densities can be explained by the fact that the assumed $L_{1.4\rm GHz}-L_{HX}$ relation \citep{damato22} is not valid for the RL AGN population, which is therefore not included in our model (see Sec.~\ref{sec:RLF}). 
To highlight the RL AGN component missed by our model, we show in Fig.~\ref{fig:RLF_smolc} the empirical RL AGN luminosity function derived by \cite{massardi10} by summing three different populations of RL AGN: flat-spectrum radio quasars, BL Lacs and a steep spectrum RL AGN population.
The three populations are described considering different evolutionary properties (see \cite{massardi10} for details). The RL AGN luminosity functions are in good agreement with the \citet{Smoilcic17c} datapoints at all luminosities (at least up to $z\sim 2.5$), and particularly at high radio luminosities, where we expect that the REX and RL AGN classifications trace the same population. \\
The excess of the modelled AGN number densities at low radio luminosities can be explained as follows. AGN with low radio luminosities are more difficult to select via the statistical AGN radio-excess selection techniques since their radio emission can be easily overwhelmed by the radio emission coming from the SF of the host galaxy. Therefore, the REX AGN sample of \citet{Smoilcic17c} is probably incomplete in this low-luminosity regime. This incompleteness almost disappears when combining multiple AGN selection techniques (see Sec.~\ref{sec:NC_comp}).

\subsection{Number counts comparison}\label{sec:NC_comp}
In Fig.~\ref{fig:number_counts} we present the Euclidean normalized differential radio NC obtained from our model compared with other models and measurements in the literature.
The green shaded area around the AGN NC,  obtained using Eq.~\ref{eq:NC}, 
reflects the assumed RLF uncertainties (see Sec.~\ref{sec:RLF}). The CTK number counts follow a trend similar to that of the total AGN population, and hence their fraction is almost constant ($\sim$40\%) with radio flux. This stems from the prescriptions of the CXB model, where $\sim 40\%$ of all AGN are assumed to be CTK (once averaged over the whole luminosity range) and from the fact that radio emission is unaffected by absorption, probing the assumed intrinsic CTK AGN fraction.\\
It is useful to compare our AGN model with the global population of radio sources, represented by the T-RECS simulated NC \citep[][gray line]{Bonaldi19}. The T-RECS simulation includes both RL AGN and SFG, dominating respectively at the highest and lowest radio fluxes.  
As discussed earlier on, our model does not include RL AGN.
We decided not to forcibly include the RL AGN population in our model for two main reasons. First, RL AGNs are expected to constitute a small fraction of the overall AGN population, $\sim 10\%$ of all the AGN, as reported by different works studying RL AGN at different redshift \citep{williams+15,liu21}. Second, they dominate the NC at $\rm S_{1.4GHz}> 0.5-1 mJy$, as it is possible to see also from Fig.~\ref{fig:number_counts} \citep[see the orange dashed line derived from the RL AGN luminosity function of ][]{massardi10}. Therefore, they are not the main focus of this paper, where we are primarily interested in quantifying the contribution of AGN at the faintest radio fluxes.
In this respect, it is worth noting that the $L_{1.4\rm GHz}-L_{HX}$ we used for our model was derived considering an X-ray selected spectroscopically confirmed AGN+ETG sample, which is mostly composed of RQ AGN, and where no pre-selection based on radio excess was done \citep{damato22}. 
This allows us to compare our models with radio-selected samples in which AGN are not only selected based on their radio excess, but also through selection criteria based on AGN signatures at other bands (like those discussed in Sec.~\ref{sec:intro}). Thus, our model is able to account for a larger population of low-luminosity radio-selected AGN. 
In Fig.~\ref{fig:number_counts} we compare our AGN model with AGN samples extracted in some of the major deep radio surveys/fields: the 3GHz JVLA-COSMOS \citep{smolcic17a,smolcic17b,Smoilcic17c}, the Extended Chandra Deep Field South (ECDFS) \citep{miller13,padovani15}, the Lockman-Hole (LH) \citep{prandoni18,bonato21} and the MIGHTEE radio survey \citep{whittam22}. 
The AGN NC shown in Fig.~\ref{fig:number_counts} as filled points are extracted from the aforementioned surveys by selecting AGN using all the available selection techniques (radio excess, X-ray emission, MIR colors, SED fitting, and spectroscopy). Instead, the squares refer to REX AGN only in the COSMOS and in the ECDFS fields. In particular, the COSMOS REX data points are derived from the same REX AGN sample that was used to build the RLF presented in Fig \ref{fig:RLF_smolc}.\\
For fluxes $\rm S_{1.4 GHz}<1 mJy$, the NC derived by selecting radio AGN using multiple criteria, are in a very good agreement with our AGN model predictions,  down to the faintest fluxes $\rm \sim 20 \mu Jy$. Instead, the data from REX AGN are below our model, with a discrepancy that increases going to fainter radio fluxes. On the contrary, the REX AGN number counts seem to be in good agreement with the RL AGN NC (orange dashed line) derived from \cite{massardi10}. This result supports the scenario discussed in Sec \ref{sec:LF_comp}: the REX AGN RLF is incomplete, in particular at low radio luminosities (and consequently low fluxes), and it mostly accounts for the brighter RL AGN population.\\
As a final remark, we notice that we made some sanity checks on our AGN model. First, we compared the NC derived from our AGN model with the NC derived by the model of \cite{lafranca10}. We found that the two radio counts do not differ significantly. In particular, the NC derived in \cite{lafranca10} predicts slightly lower AGN number densities, but still within our uncertainty region.
Second, we derived the NC considering a flatter AGN radio spectral index, namely $\alpha=0.4$, finding no significant deviation from the NC derived with $\alpha=0.7$, as again they lie within our uncertainty region.

\begin{table*}
\centering
\begin{spacing}{.1}
\begin{tblr}{colspec={M{1.7cm}M{0.7cm}M{1.6cm}M{0.7cm}M{0.1cm}M{0.7cm}M{3cm}M{0.7cm}},row{1-4}={0ex}}
\hline
\hline
\vspace{0.01cm}\\
\SetCell[r=3]{c} Field & \SetCell[c=3]{c} Radio (1.4GHz) & & &\SetCell[r=3]{c} & \SetCell[c=3]{c} X-ray (0.5-2 keV)&&\\
  \hline
     & Area & flux limit & $\rm T_{OBS}$ & & Area & flux limit & $\rm T_{OBS}$ \\
     & $\rm [deg^{-2}]$& $\rm [\mu Jy]$  & [hrs] & & $\rm [deg^{-2}]$& $\rm {\scriptsize 10^{-16}} [erg\ s^{-1}\ cm^{-2}]$ &  [hrs]\\
  \hline 
   J1030 & 0.18 & 10.6 & 30 && 0.09 & $0.57$ &140\\
    COSMOS & 2.4 & 18 & 385 && 2.15 & $1$  &1278\\
    LH-XMM & 0.31 & 22 & 75 && 0.19 & $1.9$ &322\\
    CDFS & 0.32 & 16 & 40 && 0.13 & $0.08$ &1944\\
    ECDFS & 0.32 & 16 & 40 && 0.31 & $0.74$ &69\\
    XMM-LSS & 5 & 32 & 48 && 5.36 & $2.7$ &361\\
    BOOTES & 6.7 & 80 & 192 && 9.26 & $1.5$ &944\\
    ELAIS-S1 & 2.7 & 83 & 410 && 3.2 & $4.7$ &278\\
    CDFN & 0.07 & 12 & 39 && 0.125 & $6$ &556\\
    LHN & 0.4 & 15 & 140 && 0.5 & $3.2$ &175\\
    \hline
\end{tblr}
\end{spacing}
\vspace{0.3cm}
\caption{Summary of the area covered by radio and X-ray observations, flux limits, and observing time for each of the considered fields. The X-ray and radio flux limits correspond to the flux of the faintest sources in the $0.5-2$keV X-ray catalogs and in the 1.4GHz catalogs, respectively. We note that for the CDFS and ECDFS we considered the same radio survey. The fields are listed with the references for the radio observation followed by the X-ray one: J1030 \citep{damato22,nanni20}, COSMOS \citep{smolcic17a,civano16}, LH-XMM \citep{biggs06,Brunner2008}, ECDFS \citep{miller13,xue16}, CDFS \citep{miller13,liu17}, XMM-LSS \citep{heywood20,chen18}, BOOTES \citep{deVries02,masini20}, ELAIS-S1 \citep{franzen15,ni21}, CDFN \citep{owen18,xue16}, LHN \citep[Lockman-Hole North][]{owen08,trouille08}}
\label{tab:fields}
\end{table*}

\subsection{Expectations from the main radio and X-ray deep fields}\label{sec:RX_field}
\begin{figure*}
\centering
	\includegraphics[width=1.5\columnwidth]{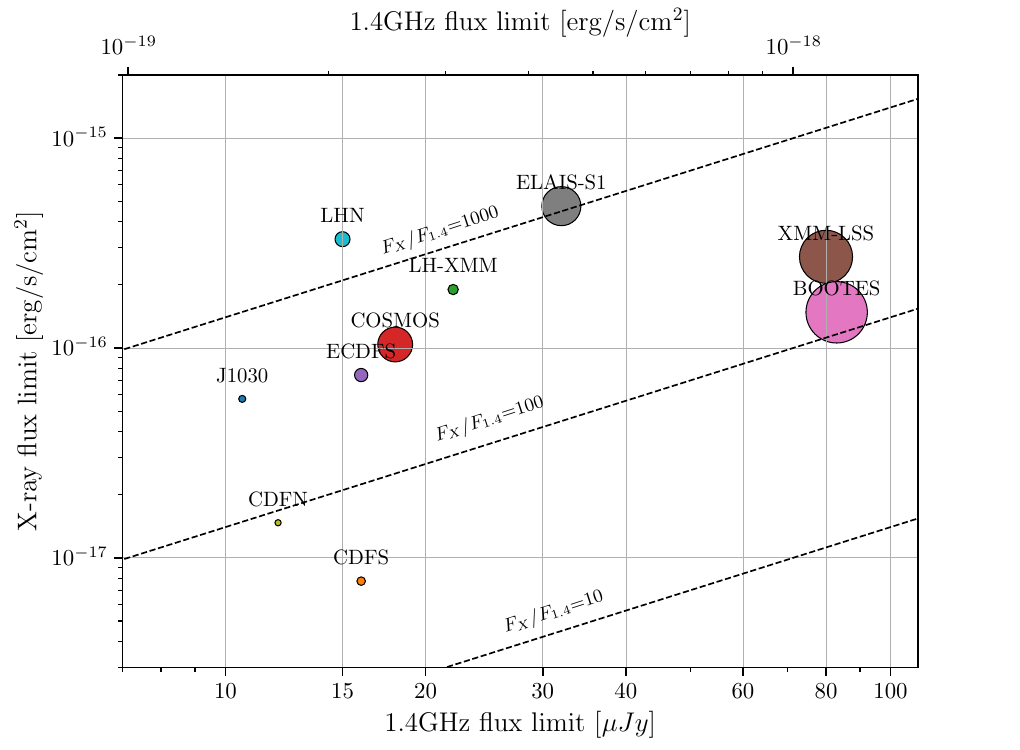}
    \caption{Distribution of the $0.5-2$keV vs 1.4GHz flux limit for the 10 radio and X-ray fields considered in this work. The reported flux limits correspond to the faintest sources' flux in the $0.5-2$keV and in the 1.4GHz catalogs, respectively. Each field marker size is proportional to the area of the overlapping region between the radio and X-ray observations}. The three dashed lines correspond to level curves for different ratios between the X-ray and radio fluxes.
    \label{fig:fields}
\end{figure*}

We are interested in assessing the relative effectiveness of deep X-ray and radio surveys to detect the elusive obscured AGN populations and particularly CTK AGN. We use the NC derived from our radio model and the X-ray population synthesis model of the CXB, to respectively predict the number of radio-detected and X-ray detected AGN and CTK AGN over some of the major deep extragalactic radio fields, also covered by X-ray observations. 
We report the areas, sensitivities, and exposure times of the considered fields in Tab.~\ref{tab:fields}. 
For the COSMOS field, the original 3GHz fluxes and limits were translated to 1.4GHz assuming a spectral shape $S_{\nu}\propto \nu^{-\alpha}$, with $\alpha=0.7$.\\
Fig.~\ref{fig:fields} shows  the X-ray soft band (SB; 0.5-2 keV) versus 1.4GHz flux limit plane for each of the 10 considered fields. 
Each field marker size is proportional to the area of the overlapping region between the observations in the two bands.

\subsubsection{Number counts corrections} \label{sec:corr_factors}

\begin{figure}
	\includegraphics[width=1\columnwidth]{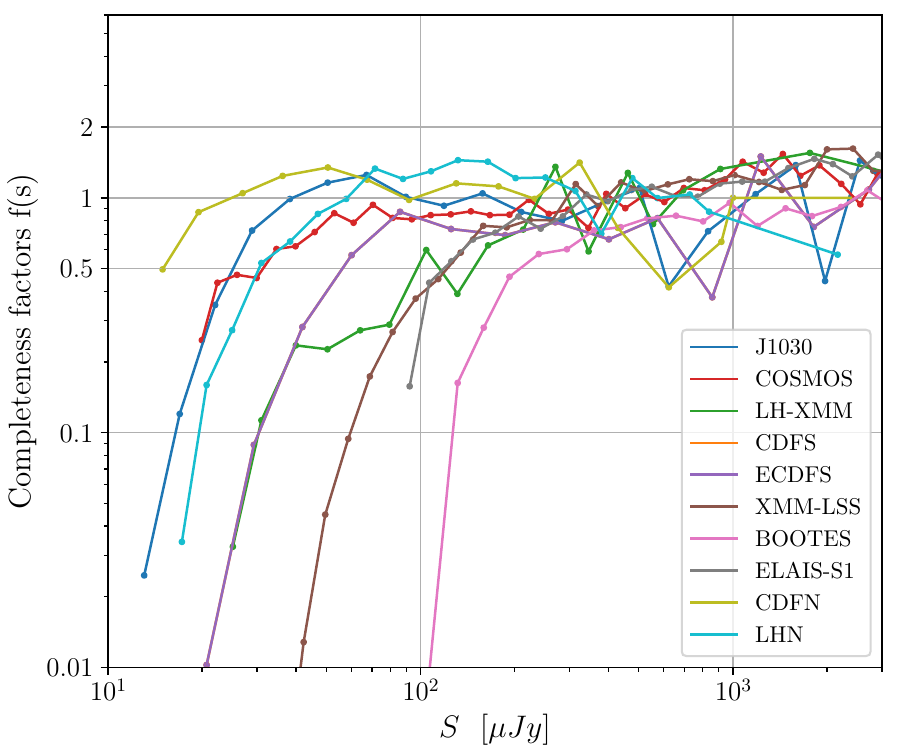}
    \caption{Completeness factors of each radio catalog derived as discussed in Sec.~\ref{sec:corr_factors}. Each color represents a different radio field, as labelled. The dots represent the center of the flux bins used to compute the differential NC corrections (see text for details).}
    \label{fig:corr_factors}
\end{figure}

Our radio model forecasts the number of AGN per $\rm deg^{-2}$ with fluxes larger than a given $\rm S_{lim}$ over an ideal radio field of constant sensitivity. 
However, in every radio pointing the sensitivity decreases from the center to its outskirt, as described by the so-called visibility function. Moreover, other issues affect the completeness of a radio catalog, such as background noise, Eddington bias \citep{eddington1913}, and resolution bias \citep{prandoni01}. The first two effects affect the incompleteness of the catalog, particularly at its faint end, while the resolution bias causes a possible loss of extended sources. To get realistic radio AGN predictions, we computed for each field  a total "completeness" function (or factor) 
that takes into account all the aforementioned effects, and we apply it to the model forecasts. In practice we proceeded as follows. For every field we first obtained the differential NC from the sources listed in the corresponding radio catalog, namely $\rm dN_{catalog}/ds$. These should be considered as 'raw' NC, as no attempt to correct for the aforementioned incompleteness effects is made in this case. Then we compared $\rm dN_{catalog}/ds$ with the differential NC expected from an ideal (fully complete) catalogue, namely $\rm dN_{ideal}/ds$.  
The ratio between these two quantities defines the completeness factor:  
\begin{equation} \label{eq:corr_factors}
    f(\Bar{s})=\frac{dN_{catalog}}{ds}(\Bar{s})\biggl{/}\frac{dN_{ideal}}{ds}(\Bar{s}),
\end{equation}
where $\Bar{s}$ is the central flux of each flux bin.
We chose the 25$\rm deg^2$ mock catalog released as part of the T-RECS simulations \citep{Bonaldi19} as ideal catalogue, as it provides a fairly good representation of the observed 1.4GHz radio source counts, both in the bright and in the faint flux regime \citep{Bonaldi19,damato22}. As mentioned in Sec.~\ref{sec:NC_comp}, the T-RECS simulations include SFG and AGN, the two main extra-galactic radio source populations detected in deep radio fields.\\
The completeness factors of the different fields considered in this work are presented in Fig.~\ref{fig:corr_factors}. The radio catalogs are largely incomplete at their faint end, while they are virtually complete ($f(\Bar{s}) \rightarrow 1$) at bright fluxes, as expected.  
The steep decrease of the completeness functions at faint fluxes is mainly driven by the visibility function, which represents the fraction of the survey area over which sources above a given flux density can be detected. Generally, faint sources can be detected only over a limited area, close to the center of the radio image, where the local noise is lowest, while the brightest sources can be detected everywhere, even at the outskirt of the radio images where the local noise is highest. This is reflected into a rapid decrease of the visibility function going to faint fluxes. In addition, the background noise introduces errors in the measurements of source fluxes that mostly affect the completeness at low signal-to-noise ratios and so mainly at low radio fluxes.\\
We caveat that the correction factors shown in Fig.~\ref{fig:corr_factors} were obtained considering both AGN and SFGs. Assuming that these corrections hold independently of the specific type of radio source considered, we can also apply them to the radio AGN-only model we built.  
This assumption can be justified by the fact that the faint AGN  detected in deep radio fields tend to occupy the same flux density regime as SFG, and tend to be unresolved, like SFG, at the typical (arcsec-scale) angular resolution of these surveys \citep[see e.g. ][]{bonzini13}. This implies that SFG and AGN are similarly affected by incompleteness effects. Also \cite{hale23}, studying the incompleteness of MIGHTEE radio images, found that the completeness corrections for the two populations are the same for a vast range of fluxes, showing only small differences for $\rm S<10\mu Jy$, a flux density limit not crossed by none of the considered radio catalogs. 
The second assumption we make is that the corrections derived in Eq.~\ref{eq:corr_factors} hold independently of the level of AGN obscuration. Since the 1.4GHz emission is unaffected by obscuration, this assumption appears to be fine.
\begin{table*}
\centering
\begin{spacing}{.3}
\begin{tblr}{colspec={M{1.7cm}M{1cm}M{1cm}M{1cm}M{1.1cm}M{1.1cm}M{0.1cm}M{1cm}M{1cm}M{1cm}M{1.1cm}M{1.1cm}},row{1-4}={0ex}}
\hline
\hline
\vspace{0.01cm}\\
\SetCell[r=3]{c} Field & &  \SetCell[c=3]{c} Radio (1.4GHz)& & & &\SetCell[r=3]{c} & & \SetCell[c=3]{c} X-ray (0.5-2 keV)& &\\
  \hline
     & \#sources & \#AGN & \#CTK & $\rm \Sigma_{AGN}$ & $\rm \Sigma_{CTK}$ & & \#sources & \#AGN & \#CTK & $\rm \Sigma_{AGN}$ & $\rm \Sigma_{CTK}$\\
     & catalog & model & model & model & model & & catalog & model & model & model & model\\
  \hline 
   J1030 & 1489 & 533 & 222 & 0.87 & 0.36 && 256 & 184 & 8 & 0.57 & 0.025\\
    COSMOS & 10830 & 4893 & 2024 & 0.57& 0.24 && 4016 &  3320& 67& 0.43& 0.01\\
    LH-XMM & 506 & 207 & 87 & 0.19 & 0.08 && 409 & 394 & 15 & 0.58 & 0.022\\
    CDFS & 883 & 361 & 149 & 0.31 & 0.13 && 1008 & 696 & 111 & 1.49 & 0.24\\
    ECDFS & 883 & 361 & 149 & 0.31 & 0.13 && 1003 & 955 & 58 & 0.86 & 0.052\\
    XMM-LSS & 5762 & 2747 & 1124 & 0.15 & 0.06 && 5242 & 6052 & 80 & 0.31 & 0.004\\
    BOOTES & 3172 & 1695 & 690 & 0.07 & 0.03 && 6891 & 5892 & 64 & 0.18 & 0.002\\
    ELAIS-S1 & 2084 & 1095 & 446 & 0.11 & 0.046 && 2630 & 2410 & 20 & 0.21 & 0.002\\
    CDFN & 795 & 270 & 112 & 1.07 & 0.44 && 683 & 545 & 66 & 1.21 & 0.147\\
    LHN & 2056 & 775 & 304 & 0.54 & 0.21 && 761 & 695 & 14 & 0.39 & 0.008\\
    \hline
\end{tblr}
\end{spacing}
\caption{Expected number of AGN (total and CTK) compared with the total number of sources observed in each field. Radio and X-ray predictions are shown in the left and right side of the table, respectively. The total and CTK AGN densities ($\Sigma$) are computed by dividing the number of predicted AGN by the area of the radio or X-ray image (see Tab.~\ref{tab:fields}) and are reported in units of $\rm arcmin^{-2}$. 1.4GHz expectations for the ECDFS and CDFS are the same since we considered the same radio survey for both.}
    \label{tab:all_pred}
\end{table*}

\subsubsection{Radio predictions}\label{sec:radio_pred}
\begin{figure}
	\includegraphics[width=1\columnwidth]{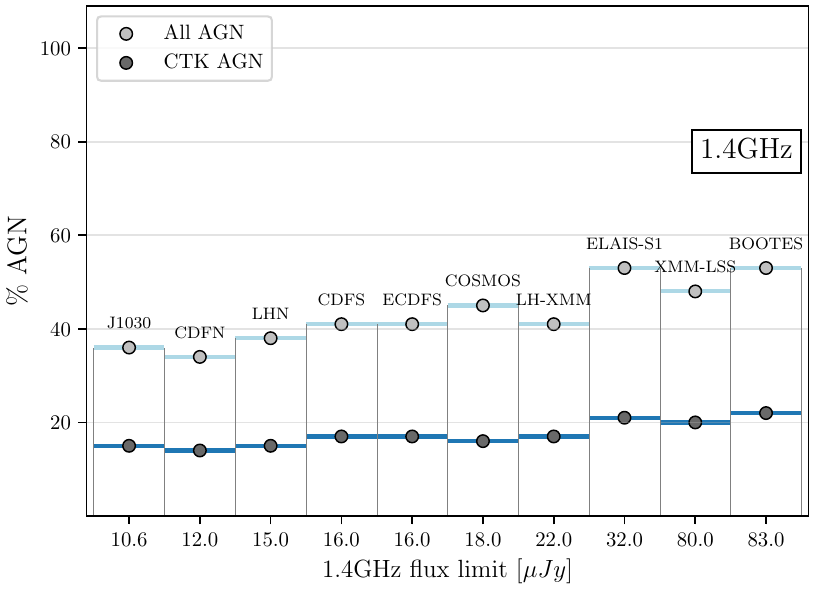}
        \vspace{10pt}
        \includegraphics[width=1\columnwidth]{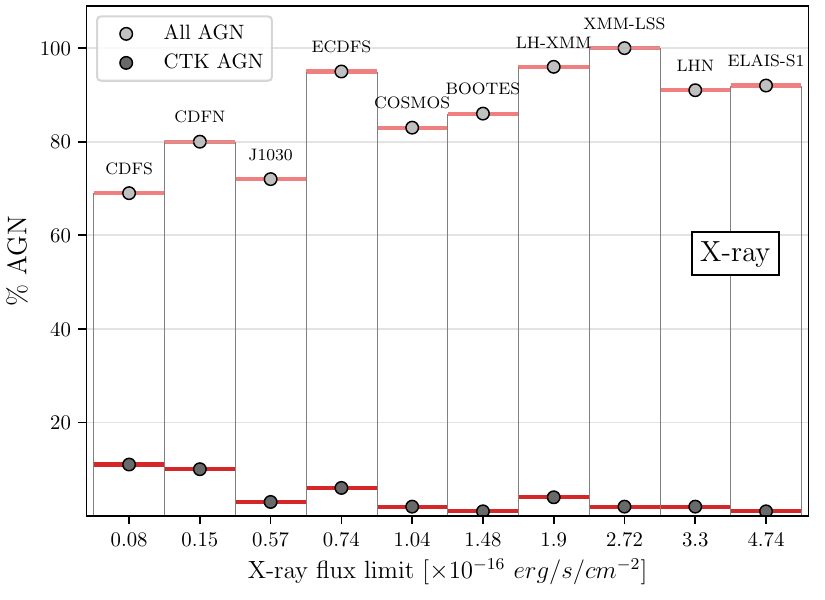}
    \caption{ {\it Upper panel:} predicted fractions of AGN (light blue) and CTK AGN (dark blue) as a function of radio survey flux limit. {\it Lower panel:} same as in the upper panel but for the X-ray predictions. In both panels the surveys are sorted on the x-axis by increasing flux limit (faintest source in the catalog). }
    \label{fig:AGN_frac}
\end{figure}

To quantify the number of AGN predicted in the different radio catalogs we applied the following equation: 
\begin{equation} \label{eq:AGN_pred_field}
   \rm N_{AGN}=A\int^{\infty}_{0} \frac{\rm dN_{AGN\ model}}{ds}(s)\cdot f(s),
\end{equation}
where $A$ is the area of the radio field, $\rm S_{min}$ is the minimum radio flux density of the radio catalog and $f(s)$ is the completeness function defined in Eq.~\ref{eq:corr_factors}.  $\rm dN_{AGN\ model}/ds$ are the differential NC derived by our model (Eq.~\ref{eq:NC}) considering $\rm\log L_{1.4 GHz\ min}=37$, $\rm\log L_{1.4 GHz\ max}=43$, $z_{min}=0$, $z_{max}=10$. 
We assigned $\rm\log L_{1.4 GHz\ min}=37$ since it is an acceptable threshold for the lower limit of the AGN 1.4GHz luminosity distribution, as reported by some of the deepest radio observations performed so far \citep{algera20,alberts20}. 
The number of AGN and CTK AGN predicted by our model at 1.4GHz for each field is presented on the left side of Tab.~\ref{tab:all_pred}.
The predictions are based on our baseline model. By using the uncertainty boundary regions described in Sec.~\ref{sec:RLF}, the prediction would get on average 30\% higher (upper boundary) or 40\%  lower (lower boundary).\\
The top panel of Fig.~\ref{fig:AGN_frac} shows the fraction of AGN and CTK AGN expected by our model obtained by dividing the predicted number of AGN by the whole number of radio sources in each radio catalog. The fields are sorted in ascending order of radio flux limit. The expected fraction of AGN ranges between 36\% in the J1030 field, the deepest radio field in our sample, up to 53\% in the BOOTES and ELAIS-S1 fields, which are among the shallowest ones. The top panel of Fig.~\ref{fig:AGN_frac} shows a clear trend where the AGN fraction in radio catalogs decreases as the sensitivity of the survey increases. This is in accordance with the evidence, presented by several works in literature and also shown in Fig.~\ref{fig:number_counts}, that AGN dominates over SFGs in the bright-fluxes regime of the global radio NCs (grey line in Fig.~\ref{fig:number_counts}).
On the contrary, the lower the radio flux, the larger the relative contribution of SFG to the global radio NC (see again Fig.~\ref{fig:number_counts}), and consequently also the fraction of AGN gets reduced.\\
All our predictions, particularly those for the shallower fields, have to be considered conservative. Indeed, the given AGN fractions do not include, by construction, the most powerful radio AGN (the RL population) that dominate the bright end of the NC.
About $\sim40\%$ of all AGN, i.e. about $\sim20\%$ of the radio sources in each catalog, are expected to be CTK nuclei, as reported also in Sec.~\ref{sec:NC_comp}.

\subsubsection{X-ray predictions}\label{sec:X_pred}
To compare radio versus X-ray AGN yields in the considered fields, we proceeded as follows. We computed the AGN differential number counts based on the CXB model, taking $\rm\log L_{X\ min}=41$, $\rm\log L_{X\ max}=47$, $z_{min}=0$, $z_{max}=10$. The X-ray lower limit $\rm\log L_{X\ min}=41$ corresponds to the radio lower limit $\rm\log L_{1.4 GHz\ min}=37$ used in Eq.~\ref{eq:AGN_pred_field}, when assuming  the $L_{1.4\rm GHz}-L_{HX}$ relation of our model. This value is also a reasonable lower limit to the known AGN X-ray luminosity distribution, and we checked that this choice does not violate the total CXB flux constraint. Using the sky coverage $A(s)$ and the CXB differential NC, $\rm dN_{AGN\ CXB}/ds$, the number of AGN over a certain X-ray field is predicted as follows:
\begin{equation} \label{eq:AGN_pred_field_x}
    \rm N_{AGN}=\int^{\infty}_{0} \frac{\rm dN_{AGN\ CXB}}{ds}(s)\cdot A(s),
\end{equation}
where $S_{min}$ is the minimum SB X-ray flux of the respective catalog and $A(s)$  represents the sky coverage, namely the observed area as a function of the sensitivity of the survey. The function $A(s)$ is generally retrieved by performing completeness simulations, i.e. already implementing any possible completeness correction factor.
The results are reported on the right side of Tab.~\ref{tab:all_pred}.\\
We show only the predictions in the SB for two reasons. Firstly, the numbers of AGN and CTK AGN predicted by the CXB model in the SB are larger than in the hard band (HB, $2-10$keV), for every field. Indeed, for all the considered fields, the flux limit in the SB is always lower by a factor$\sim 5-10$ than the one in the HB, according to the fact that for all the X-ray catalogs the number of sources detected in the SB is larger than in the HB\footnote{The only exception is the J1030 field, where the number of X-ray sources reported in \cite{nanni20} in the SB (HB) is 193 (208). However, the Chandra observation of the J1030 filed is the latest one among those considered in this work and suffers from the SB Chandra sensitivity deterioration  \citep{peca21}}. Secondly, the expected fraction of CTK AGN predicted by the CXB model is larger in the SB than in the HB for $z>0$. This happens because of the deeper SB flux limit and because of the presence of a minor (but not negligible) SB reflected component in the CTK AGN rest-frame spectrum implemented in the CXB model \citep[Fig.1 in ][]{gilli07}. 
We finally notice that in the case of the XMM-LSS field, the number of AGN predicted by the X-ray model exceeds the number of X-ray sources in the related catalog by a fraction $\sim15\%$. This excess can be justified by field-to-field variation of the number counts or by an overestimate of the CXB model in the SB already observed in a previous work \citep{marchesi20}.\\
In the lower panel of Fig.~\ref{fig:AGN_frac} we report the fraction of X-ray AGN and CTK AGN predicted over the different X-ray fields sorted by increasing X-ray flux limit. In the X-ray band AGN are the dominant population: in all cases the fraction of AGN is $\geq70\%$, with again an increasing trend for shallower flux limits. Indeed, the X-ray emission of SFGs starts to be detected only at very faint fluxes, and are rarely detected in the shallower surveys. The increasing trend of the AGN fraction going to higher flux limits in the X-ray panel is more scattered than in the radio panel. This is mainly due to the shapes of the sky coverage of the different X-ray observations that can change significantly from field to field.
In general, X-ray surveys allow the detection of a larger fraction of AGN with respect to radio ones, where the contamination from SFG can be large,  especially at faint fluxes.
However, the large majority of the AGN detected in the X-rays are not heavily obscured: the fraction of CTK AGN in the X-rays is always between 1\%-11\% of all the X-ray sources and between the 1\%-15\% of all the X-ray AGN predicted by the CXB model, the larger fraction reached in the deepest survey, namely the CDFS. Indeed, the fraction of detectable CTK AGN shows a remarkable dependence on the X-ray flux limit: the lower the X-ray flux limit the higher the fraction of CTK AGN detected. This trend is justified by the fact that heavily obscured sources are usually very faint in the X-rays and so their detection requires very deep surveys. On the contrary, in the radio band, the CTK AGN are between $14-22\%$ of all the radio sources and $\sim 40\%$ of all the radio AGN predicted by our model, and show an opposite dependence on radio flux limit. 

\subsection{AGN densities}\label{sec:AGN_density}
\begin{figure}
	\includegraphics[width=1\columnwidth]{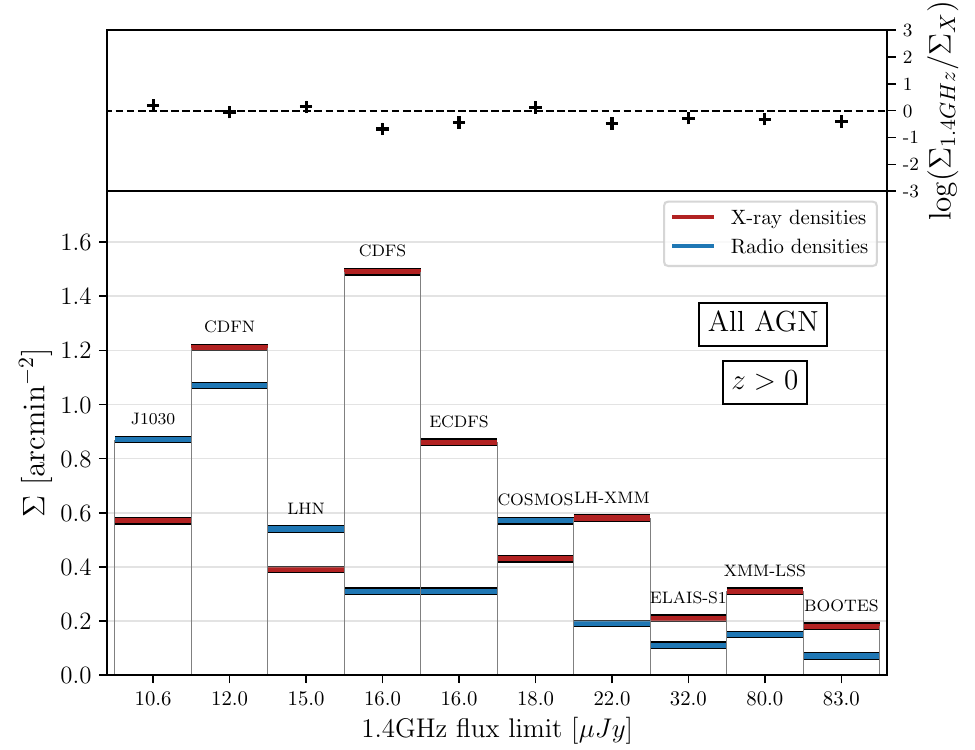}
        \vspace{1cm}
        \includegraphics[width=1\columnwidth]{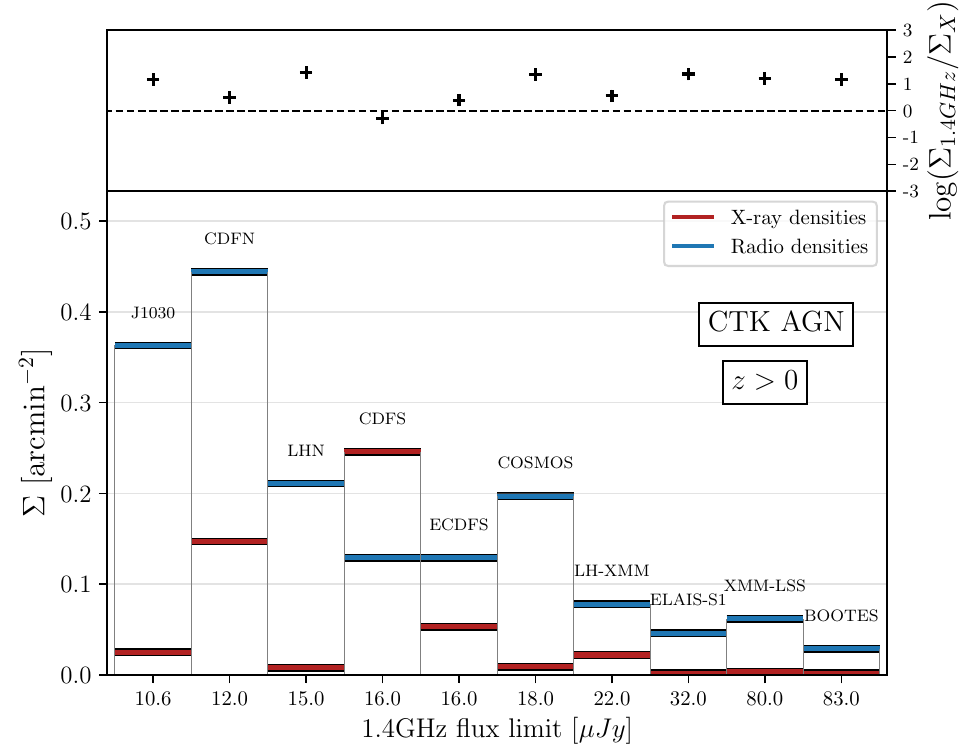}
    \caption{{\it Upper panel:} AGN densities in X-ray (red) and radio (blue) surveys as a function of 1.4GHz survey flux limit, as reported in Tab.~\ref{tab:all_pred}. The ratio between the 1.4GHz and X-ray AGN densities is shown in the top position of the plot. {\it Lower panel:} same as in the upper panel but for CTK AGN. In both panels the surveys are sorted on the x-axis by increasing radio flux limit (faintest source in the catalog).}
    \label{fig:AGN_density}
\end{figure}
In Fig.~\ref{fig:AGN_density} we plot the radio and X-ray densities predicted by the two models for both total and CTK AGN. These densities, also reported in Tab.~\ref{tab:all_pred}, are computed by dividing the number of AGN predicted by the area of the radio or X-ray observations.
In the upper panel, we note that the X-ray observations generally return a larger density of AGN per $\rm arcmin^{-2}$, except for the J1030, LHN and COSMOS fields. However, the differences between the densities predicted in the two bands are generally less than one order of magnitude. When only the CTK population is taken into account (lower panel) the situation is the opposite: for every field (except the CDFS) the CTK AGN density is much larger in the radio than in the X-rays, by even more than one order of magnitude. Furthermore, as one would expect, the 1.4GHz AGN and CTK AGN densities increase for deeper radio observations.\\
These two plots clearly show that X-ray observations are on average more effective than radio ones in detecting AGN when the global AGN population is considered, but if we want to detect the most obscured AGN the radio selection is more effective than the X-ray one (by on average a factor of 10 in term of surface densities). In Sec.~\ref{sec:hz} we show that this result holds even at high redshift.

\section{Discussion} \label{sec:Discussion}

\subsection{Radio catalogs: AGN predictions vs observations} \label{sec:model_vs_obs}
Since the radio model we built is not set to predict only the REX AGN population but, as presented in Sec.~\ref{sec:NC_comp}, it can be used to forecast the whole population of radio-quiet AGN we compared our results with those found in the literature using multiple selection techniques.\\
In the case of the COSMOS field, as mentioned in Sec.~\ref{sec:LF_comp}, \cite{smolcic17b} distinguished radio sources into pure SFG, medium-luminosity AGN (MLAGN), and high-luminosity AGN (HLAGN). Using color selection, X-ray counterpart analysis, SED fitting decomposition, and radio excess, they identified 1623 HLAGN and 1648 MLAGN among all the radio sources with a counterpart in the COSMOS2015 photometric catalog \citep{laigle16}. All the AGNs represent around the 42\% of the COSMOS radio sources in the reference catalog, a fraction very close to the 45\% of AGN our model predicts for COSMOS. However, our predictions have to be considered conservative since our model does not take into account the classical RL AGN, which are expected to be around 10\% of the whole AGN population. In the same work, \cite{smolcic17b} made also an AGN selection based only on the radio information, selecting only those AGN showing a ($>3\sigma$) radio excess with respect to what is expected from a pure SFG, to ensure that at least the 80\% of their radio emission is due to the AGN component. Their final REX AGN sample is constituted by 1846 sources, i.e. by only 26\% of all the radio sources in COSMOS. This shows that the radio excess technique alone is not able to provide complete AGN samples, and multiple selection methods are needed to this end.\\
Recent MIGHTEE observations of the COSMOS field \citep{whittam22,zhu23} obtained results similar to those of \cite{smolcic17b}. The 1.4GHz MIGHTEE data in the COSMOS field reach almost the same sensitivity as the 3GHz VLA COSMOS image, but was obtained on a smaller portion of the field $\rm \sim 1.6 deg^2$. Using the radio excess criterion, following the more recent prescriptions of \cite{delvecchio21}, \cite{whittam22} found 1332 radio excess AGN over a sample of 5223 sources, corresponding to  $\sim 25\%$. On the contrary, when they account for all the other selection techniques, they retrieve a fraction of AGN around 35\%. \cite{whittam22} also present a consistent fraction of unclassified sources; if these are split into AGN and SFG based on the flux density ratio of the classified sources, the fraction of the whole AGN gets to $\sim 40\%$, close to our predictions.\\
Another example comes from the radio observation of the CDFS \citep{miller13}, where source classification is reported in \cite{bonzini13} and \cite{padovani15}. Among the 883 radio sources in their catalog, they identified 381 AGN (43\%) using IRAC colors, X-ray counterparts, and the radio excess parameter. Our model predicts a similar number of AGN (361) that has to be taken as a conservative estimation since RL AGN are excluded. \cite{bonzini13} selected as REX AGN only 173 sources, only 19\% of the whole radio sources, again suggesting that when multiple, deep, multiband analyses are adopted the completeness of the radio AGN sample largely increases and the number of AGN get very similar to what is predicted by our model.\\
The radio selection of heavily obscured AGN has been only marginally investigated in the literature. The works mentioned above did not investigate the radio properties of AGN in terms of their obscuration levels. Instead \cite{andonie22}, starting from an IR-based AGN selection in the COSMOS field (and limiting to $z<3$), investigated the obscuration properties of their sample by means of X-ray and radio data. They found that 73\% of IR-selected AGN were in the VLA 1.4GHz or 3GHz catalogs, and in particular that 63\% of these radio detected AGN are obscured ($\rm \log N_{H}>22$).\\ 
Despite the broad agreement with the total number of AGN observed in radio catalogs, we remark that the number of heavily obscured CTK AGN expected by our model is driven by the assumptions made in the CXB modeling, and is therefore affected by the uncertainties discussed below.
The population synthesis model of the CXB considers X-ray data that extend beyond the traditional 0.5-10 keV band (the peak of the CXB is around 20-30keV), allowing one to constrain the population of heavily obscured AGN better than what is possible with Chandra and XMM observations alone. 
However, the derived total abundance of CTK AGN, and in particular that of the most heavily obscured, reflection-dominated AGN ($\rm \log N_{H}>25$), is degenerate with the reflection efficiency, and hence overall normalization, assumed for their X-ray spectra.
In principle, by decreasing the reflection efficiency by a factor of $\sim2$, which would still be consistent with broad band X-ray observations of local CTK AGN, would increase
the number of reflection-dominated CTK AGN by an equal factor. An additional uncertainty is related to the scatter of the measured CXB intensities. For example, \cite{comastri15} showed that by assuming a space density of reflection-dominated CTK AGN 4x larger than what was assumed by \cite{gilli07}, the produced CXB flux would still be in agreement with the CXB measurements within their scatter. However, we do not expect the fraction of CTK AGN to be severely higher (or lower) than what we assumed, as suggested by number of works in the literature that searched for CTK AGN by means of multi-band tracers that are in principle not affected by obscuration and can be used as proxies of the intrinsic AGN nuclear power. For example, \cite{daddi07} and \cite{fiore09} selected CTK AGN by means of their mid-IR excess (hot dust) emission as compared to their X-ray emission, finding space densities of CTK AGN at $z\sim1.5-2.0$ consistent with the CXB model expectations. A similar agreement was found at $z\sim0.8$ by \cite{vignali14} who selected obscured AGN through their narrow [Ne v]3426 Å emission line. The uncertainties related to the high-redshift predictions of our model will be discussed in Sec. \ref{sec:hz}.

\subsection{AGN detection: Radio or X-ray?}\label{sec:radio_X?}

\begin{figure}
	\includegraphics[width=1.\columnwidth]{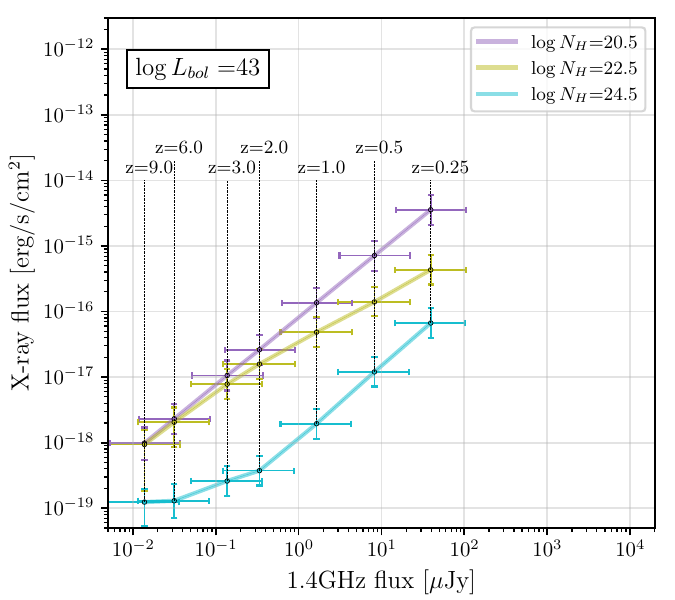}
        \includegraphics[width=1.\columnwidth]{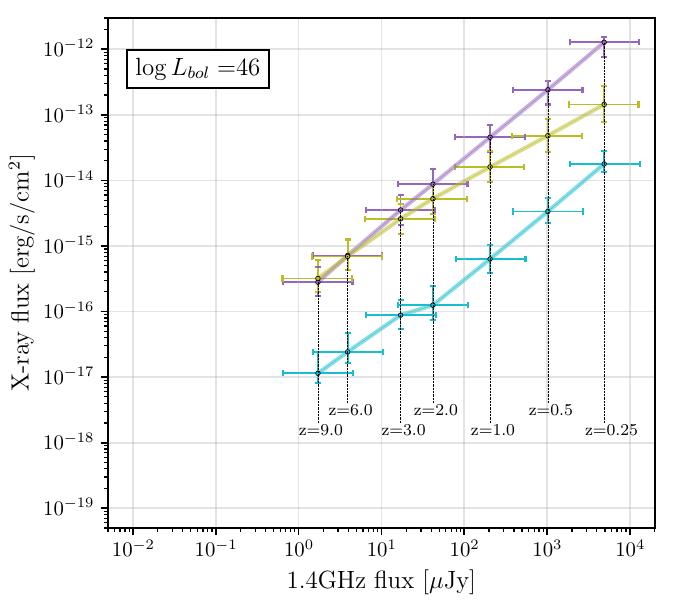}
    \caption{Typical SB X-ray flux versus 1.4GHz flux at different redshift computed for an AGN with $\rm \log L_{bol}=43$ (upper panel) and $\rm \log L_{bol}=46$ (lower panel). Error bars show the $1\sigma$ distribution values due to the scatter in the $\rm L_{bol}-L_{HX}$ relation (vertical error bar), and in the $L_{1.4\rm GHz}-L_{HX}$ relation (horizontal error bar). The dashed lines indicate the redshift at which the fluxes are computed.}
    \label{fig:det_fluxes}
\end{figure}

\begin{figure*}
	\includegraphics[width=1.\columnwidth]{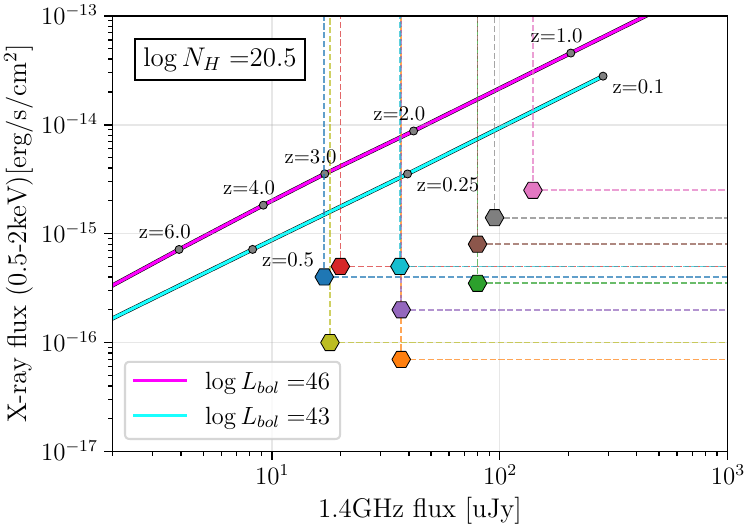}
        \includegraphics[width=1.\columnwidth]{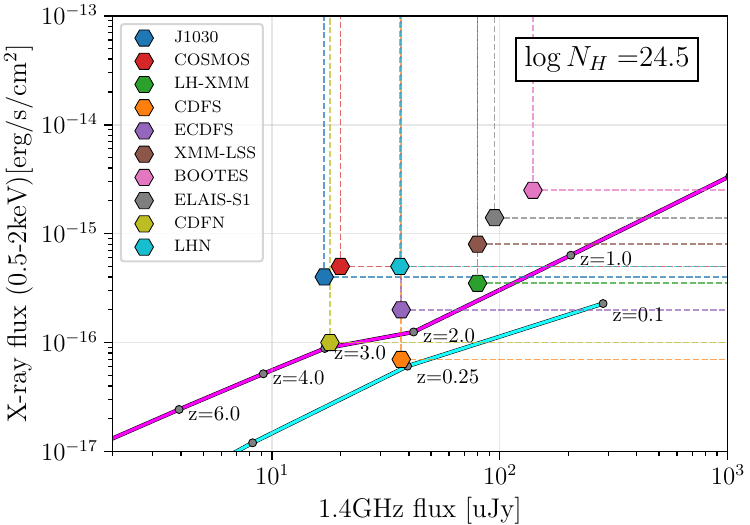}
    \caption{Median SB X-ray and 1.4GHz detection fluxes for each of the fields color-coded as labelled in the right panel. The two panels consider tracks for two different levels of AGN obscuration: $\log N_H=20.5$ (left) and $\log N_H=24.5$ (right). In cyan and in magenta we show the typical $0.5-2$keV versus 1.4GHz flux tracks, computed at different redshifts, for an AGN with $\rm \log L_{bol}=43$ and $\rm \log L_{bol}=46$, respectively. }
    \label{fig:det_limit}
\end{figure*}

For a given combination of radio and X-ray flux limits the relative efficiency of the two bands in detecting an AGN depends on the shape of the AGN spectral energy distribution, which in turn depends on a number of physical properties like the SMBH's accretion parameters, the presence of radio jets, the duty cycle, the bolometric luminosity and the obscuration level. \\
Here we want to investigate the detection efficiency of 1.4GHz and X-ray selections by varying three main AGN parameters: redshift, bolometric luminosity and obscuration level.
Starting from a fixed bolometric luminosity, we computed the HB X-ray AGN luminosity using the bolometric correction derived in \cite{duras20}, which are valid both for type 1 and type 2 AGN. Then we converted the HB X-ray luminosity into the corresponding SB luminosity assuming a power-law spectrum with photon index $\Gamma=1.9$ \citep{piconcelli05,gilli07}. To derive the X-ray flux we used one of the X-ray source mock catalogs presented in \cite{marchesi20}. The considered catalog contains 5.4M sources simulated down to very faint X-ray fluxes, $\rm 10^{-20} erg\ cm^{-2}\ s^{-1}$, over an area of 100$\rm deg^2$ and with different obscuring hydrogen column densities, following the prescription of the CXB model in G07. The catalog provides the SB flux and the corresponding SB X-ray luminosity for each source at a given redshift and obscuration level. Therefore, from given combination of SB X-ray luminosity and obscuration, we derived X-ray flux versus redshift dependences.\\
The intrinsic X-ray luminosity was then converted into 1.4GHz luminosity using Eq.~\ref{eq:LR_LX}. Finally, by considering the 1.4GHz luminosity-flux relation:
\begin{equation}
\rm \frac{L_{1.4GHz}}{\nu_{1.4GHz}}= 4\pi D_L^2\frac{S_{1.4GHz}}{(1+z)^{1-\alpha}}
\end{equation}
we obtain the corresponding radio flux density ($\rm S_{1.4GHz}$).
In Fig \ref{fig:det_fluxes} we report the AGN X-ray versus radio flux density for redshifts between $z=0$ and $z=9$, considering two different bolometric luminosities and three different levels of AGN obscuration. We also report as error bars the 1$\sigma$ intervals produced by the scatter in the $\rm L_{bol}-L_{HX}$ relation ($\sigma=0.27$ dex, vertical direction), and in the $L_{1.4\rm GHz}-L_{HX}$ relation ($\sigma=0.5$ dex, horizontal direction).\\
At any given redshift, the 1.4GHz AGN flux density is the same for the three levels of AGN obscuration, since the radio emission is not affected by obscuration. On the contrary, the X-ray fluxes at the same redshift differ significantly and decrease with increasing obscuration. We notice that the X-ray flux of an obscured AGN ($\rm \log N_H=22.5$) at a given bolometric luminosity tends to approximate the flux of unobscured AGN ($\rm \log N_H=20.5$) for $z>3$. This is due to the effect of the K-correction that shifts absorption out of the X-ray bandpass for AGN at high-z. 
Considering the two values of $\rm \log L_{bol}$ the trend of the AGN fluxes are similar for corresponding values of $N_H$, but with a shift of $\sim$2 orders of magnitude in radio and of 2.5 dex in the X-ray. \\

To understand which AGN are detectable in the radio and/or X-ray fields considered in this work, we compared these curves with the fields' X-ray and radio median sensitivities. In each panel of Fig.~\ref{fig:det_limit} we plot the 1.4GHz median flux limit (5$\sigma_{med}$) versus the X-ray sensitivity at 50\% of the field area (that can be seen as a median X-ray flux limit).
These are compared with the tracks showing X-ray versus radio fluxes for AGN at two given obscuration levels and  two different bolometric luminosities.\\ 
As shown in Fig \ref{fig:det_limit}, the tracks for unobscured AGN ($\rm \log N_H=20.5$; left panel), intercept the vertical projection of all the fields' positions before intersecting the horizontal ones. This means that, when typical unobscured AGN reaches the median radio sensitivity of the fields, its X-ray flux is largely above the median X-ray sensitivity of the same fields. Therefore, when an unobscured AGN is detected in the radio band, it is already detected also in the X-ray image of the same field.
This justifies why, as we see in Sec.~\ref{sec:X_pred}, unobscured AGN are preferentially detected in the X-rays for all the considered fields.\\
A very different behaviour is observed for CTK AGN, with $\rm\log N_H=24.5$ (right panel). In this case, for most of the fields, the AGN tracks intercept the horizontal projections before the vertical ones, meaning that when one of these AGN becomes detectable in the X-ray it is already detected in the radio image of the field. The only field for which this is not true for both values of $\rm \log L_{bol}$ is the 7Ms CDFS, where, due to the deep Chandra imaging, X-ray selection is more effective than radio selection in identifying AGN even in the presence of heavy obscuration. 
In the 2Ms CDFN, the detection efficiency of luminous ($\rm \log L_{bol}=46$) CTK AGN is the same for both radio and X-ray observations: the tracks in Fig.~\ref{fig:det_limit} (right) exactly crosses the CDFN datapoints. Less luminous CTK AGN are instead preferentially detected in the radio band.

\subsection{High redshift predictions}\label{sec:hz}

\begin{figure}
	\includegraphics[width=1\columnwidth]{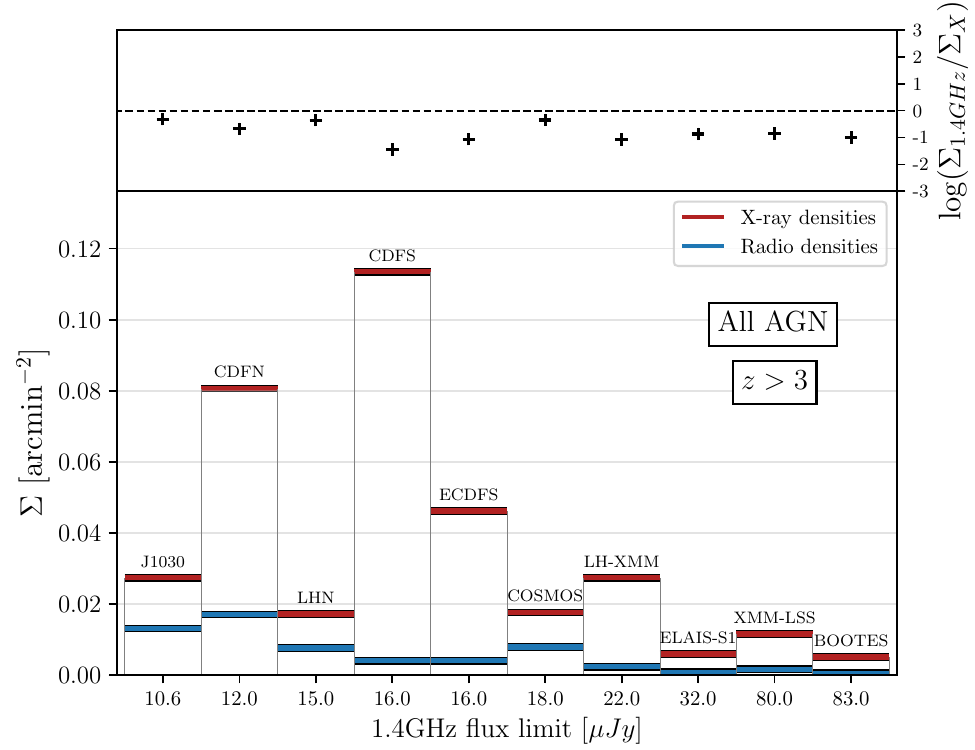}
        \vspace{1cm}
        \includegraphics[width=1\columnwidth]{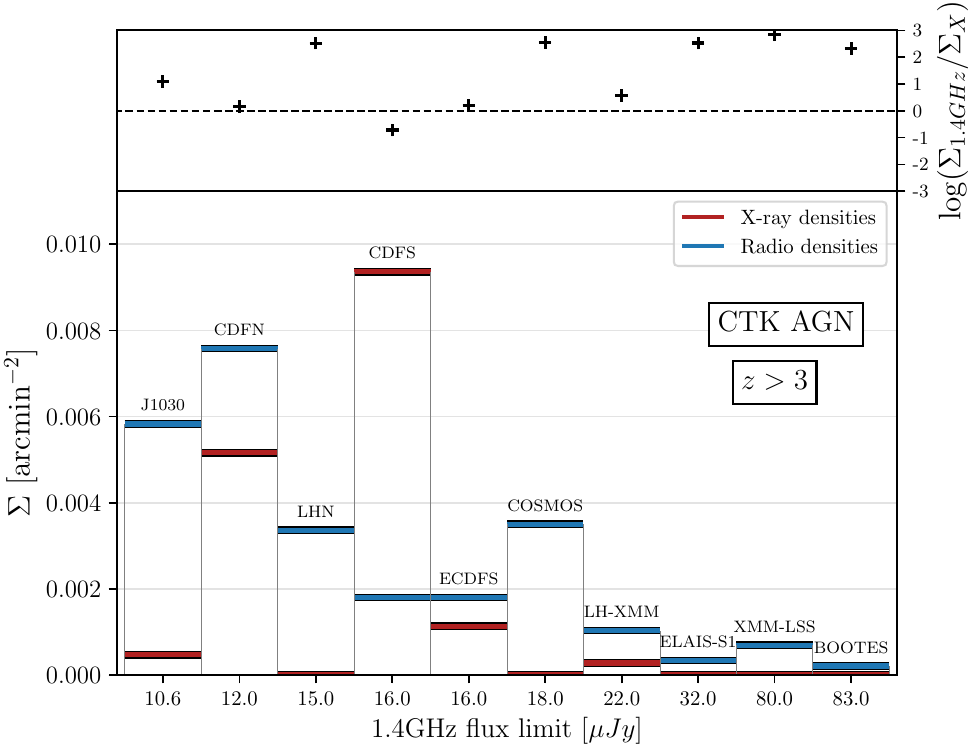}
    \caption{Same as Fig.~\ref{fig:AGN_density}, but for $z>3$.}
    \label{fig:hz_AGN_dens}
\end{figure}

\begin{table*}
\centering
\begin{spacing}{0.3}
\begin{tblr}{colspec={M{1.7cm}M{1cm}M{1cm}M{1.3cm}M{1.3cm}M{0.1cm}M{1cm}M{1cm}M{1.3cm}M{1.3cm}},row{1-4}={0ex}}
\hline
\hline
\SetCell[c=10]{c} Results $z>3$  & & & & & & & & & 
\vspace{0.01cm}\\
\hline
\vspace{0.01cm}\\
\SetCell[r=3]{c} Field & &  \SetCell[c=3]{c} Radio (1.4GHz) & & &\SetCell[r=3]{c} & & \SetCell[c=3]{c} X-ray (0.5-2 keV)& &\\
  \hline
& \#AGN & \#CTK & $\rm \Sigma_{AGN}$ & $\rm \Sigma_{CTK}$ & & \#AGN & \#CTK & $\rm \Sigma_{AGN}$ & $\rm \Sigma_{CTK}$\\
&model & model & model & model & & model & model & model & model\\
  \hline 
   J1030 & 8.0 & 3.5 & 0.013 & 5.8$\times 10^{-3}$ && 8.8 & 0.15 & 0.027 & 4.7$\times 10^{-4}$\\
    COSMOS & 68 & 30 & 0.008 & 0.35$\times 10^{-3}$ &&  135 & 0.06 & 0.017& 0.1$\times 10^{-4}$\\
    LH-XMM & 2.6 & 1.1 & 0.002 & 1$\times 10^{-3}$ && 19 & 0.19 & 0.027 & 2.7$\times 10^{-4}$\\
    CDFS &  4.6 & 2 & 0.004 & 1.8$\times 10^{-3}$ &&  53 & 4.4 & 0.113 & 9.3$\times 10^{-3}$\\
    ECDFS & 4.6 & 2 & 0.004 & 1.8$\times 10^{-3}$ &&  51 & 1.3 & 0.046 & 1.1$\times 10^{-3}$\\
    XMM-LSS & 28 & 12 & 0.002 & 7$\times 10^{-4}$ &&  222 & 0.005 & 0.011 & 1$\times 10^{-5}$\\
    BOOTES & 11 & 5.1 & 0.0005 & 2$\times 10^{-4}$ &&  168 & 0.11 & 0.005 & 1$\times 10^{-5}$\\
    ELAIS-S1 & 7.4 & 3.3 & 0.0008 & 3$\times 10^{-4}$ &&  68 & 1$\times 10^{-4}$ & 0.006 & 1$\times 10^{-5}$\\
    CDFN & 4.3 & 1.9 & 0.017 & 7.6$\times 10^{-3}$ &&  36 & 2.3 & 0.08 & 5.2$\times 10^{-3}$\\
    LHN &  11 & 4.8 & 0.008 & 3.3$\times 10^{-3}$ &&  31 & 0.01 & 0.017 & 1$\times 10^{-5}$\\
    \hline
\end{tblr}
\end{spacing}
\vspace{0.3cm}
\caption{Expected number and surface densities of $z>3$ AGN (total and CTK) in each field both in the radio (left) and X-ray band (right).}
    \label{tab:hz_pred}
\end{table*}
In Tab.~\ref{tab:hz_pred} we report the predictions of our model for AGN and CTK AGN at $z>3$. 
For consistency with the assumptions in our baseline model for radio AGN, we will use for the X-rays the expectations of the \cite{vito14} as implemented in the corresponding mock catalog \citep[see ][]{marchesi20}.
The procedure followed to forecast the number of detectable high-z AGN is the same as reported in Sec.~\ref{sec:radio_pred} and Sec.~\ref{sec:X_pred}. \\
The change in the starting HXLF also allows us to introduce an intrinsic increment in the fraction of heavily obscured AGN at $z>3$, that has been observed or predicted by different works in the literature \citep{aird15, ananna19, gilli22,ni20,lapi20}. \\
In the CXB model the fraction of CTK AGN depends on the luminosity and reaches 4/9 only for the low luminosity regime or, equivalently, at the faintest X-ray fluxes \citep[see Eq. 4 in ][]{gilli07}. On the contrary, in \cite{vito14} HXLF, we assumed \citep[according to ][]{vito18, marchesi20} a luminosity-independent and constant fraction of CTK AGN equal to 4/9 of the whole AGN population. \\
The number of X-ray AGN reported in Tab.~\ref{tab:hz_pred} are obtained considering the $0.5-2\rm keV$ X-ray band, for which the total number of AGN detected at $z>3$ is larger than in the HB. We verified that using the HB, we would have obtained larger numbers of CTK AGN (within a factor of 2), but we kept the SB information for consistency with the previous sections.\\ 
In Fig.~\ref{fig:hz_AGN_dens} we show the densities of AGN (upper panel) and CTK AGN (lower panel) both in the radio and X-ray band. The densities of $z>3$ AGN are, for both the bands, $\sim 1.5$ dex lower than what was predicted for $z>0$, remarking the difficulty of detecting such high-redshift sources with surveys with the current facilities.
Even at $z>3$, X-ray observations are more efficient in detecting unobscured or mildly obscured AGN (predicting AGN surface densities on average 10 times larger than the radio ones), whereas the radio model predicts much larger CTK AGN densities for all the fields (except the CDFS). At $z>3$ the surface density of CTK AGN radio detected is generally more than 10 times larger than the X-ray one and more than $100$ times larger for half of the fields. This strongly supports the radio selection's effectiveness in detecting heavily obscured AGN, even at high redshift. 

\subsection{SKA predictions}\label{sec:SKA_pred}

\begin{table*}
\centering
\begin{spacing}{0.8}
\begin{tblr}{colspec={M{1.6cm}M{0.7cm}M{2.3cm}M{0.75cm}M{0.75cm}M{0.75cm}M{0.01cm}M{0.75cm}M{0.75cm}M{0.75cm}M{0.01cm}M{0.75cm}M{0.75cm}M{0.75cm}},row{1-4}={0ex}}
\hline
\hline\\
\SetCell[c=14]{c} SKA high-z predictions  & & & & & &\\
\vspace{0.01cm}\\
\hline
\SetCell[r=2]{c} Survey & \SetCell[r=2]{c}Area &\SetCell[r=2]{c} Sensitivity (5$\sigma$) & \SetCell[c=3]{c} $z>3$ && &\SetCell[r=2]{c}& \SetCell[c=3]{c}  $z>6$ &&&\SetCell[r=2]{c}& \SetCell[c=3]{c} $z>10$ \\
    \hline
   &&& \#AGN & \#CTK & logL$_{\rm bol}$ && \#AGN & \#CTK & logL$_{\rm bol}$&& \#AGN & \#CTK & logL$_{\rm bol}$ \\
    \hline
    \SetCell[r=2]{c} Ultra-Deep &\SetCell[r=2]{c}  1 &\SetCell[r=2]{c} 0.25 &\SetCell[r=2]{c} 1870 &\SetCell[r=2]{c} 831 &\SetCell[r=2]{c} 43.8 &&\SetCell[r=2]{c} 34 &\SetCell[r=2]{c} 15 &\SetCell[r=2]{c} 44.4 &&\SetCell[r=2]{c} 2 &\SetCell[r=2]{c} 1 &\SetCell[r=2]{c} 44.8 \\
    &&&&&&&&&&&&\\
      \hline
    \SetCell[r=2]{c}Deep &\SetCell[r=2]{c} 20 &\SetCell[r=2]{c} 1 &\SetCell[r=2]{c} 13780 &\SetCell[r=2]{c} 6120 &\SetCell[r=2]{c} 44.4 &&\SetCell[r=2]{c}  220 &\SetCell[r=2]{c} 98 &\SetCell[r=2]{c} 44.9 &&\SetCell[r=2]{c} 6 &\SetCell[r=2]{c} 3 &\SetCell[r=2]{c} 45.6\\
    &&&&&&&&&&&&\\
      \hline
    \SetCell[r=2]{c} Wide & \SetCell[r=2]{c} 1000 &\SetCell[r=2]{c} 5 &\SetCell[r=2]{c} 182000 &\SetCell[r=2]{c} 81000  &\SetCell[r=2]{c} 45.1 && \SetCell[r=2]{c}1980 &\SetCell[r=2]{c} 880&\SetCell[r=2]{c} 45.4&&\SetCell[r=2]{c}35&\SetCell[r=2]{c} 16&\SetCell[r=2]{c}46.6\\
    &&&&&&&&&&&&\\
      \hline 
    \hline
\end{tblr}
\end{spacing}
\vspace{0.3cm}
\caption{Expected number of high-z AGN in the three 1.4GHz continuum surveys planned with SKAO. Area and sensitivities of the three tiers are reported in unit of deg$^2$ and $\mu Jy$, respectively. The values of $\log L_{bol}$ refer to the median values in the simulated samples described in Sec.~\ref{sec:SKA_pred}.}
    \label{tab:SKA_pred_v2}
\end{table*}

\begin{figure*}
\centering
        \includegraphics[width=2\columnwidth]{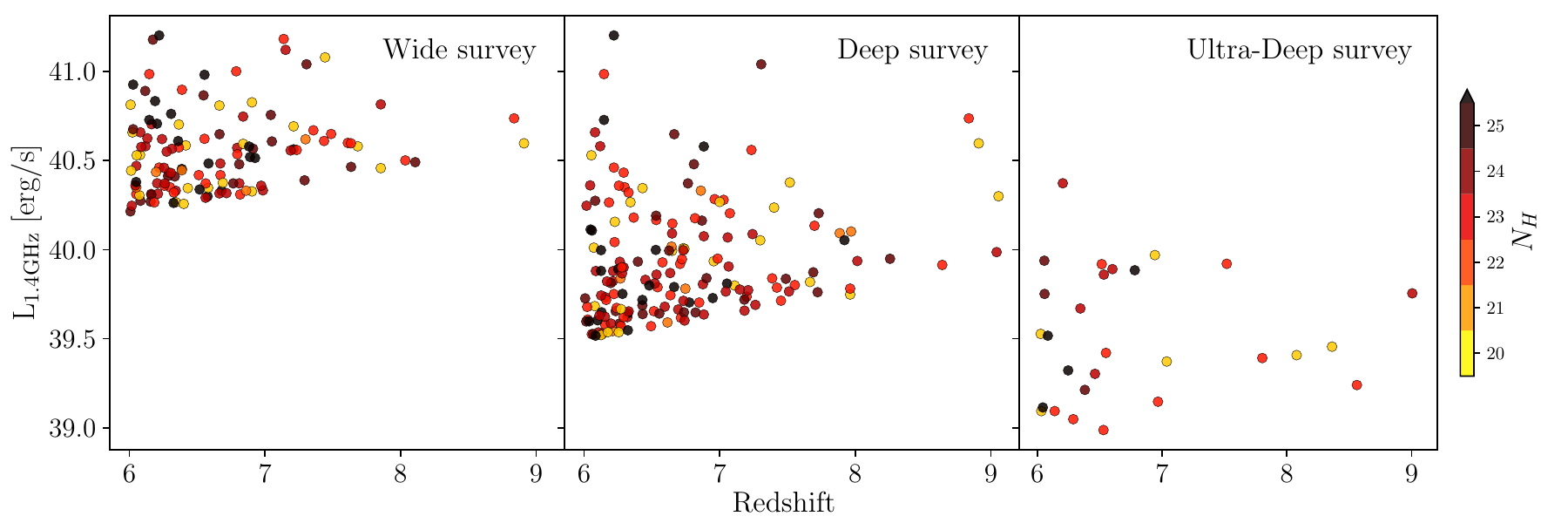}
        \includegraphics[width=2\columnwidth]{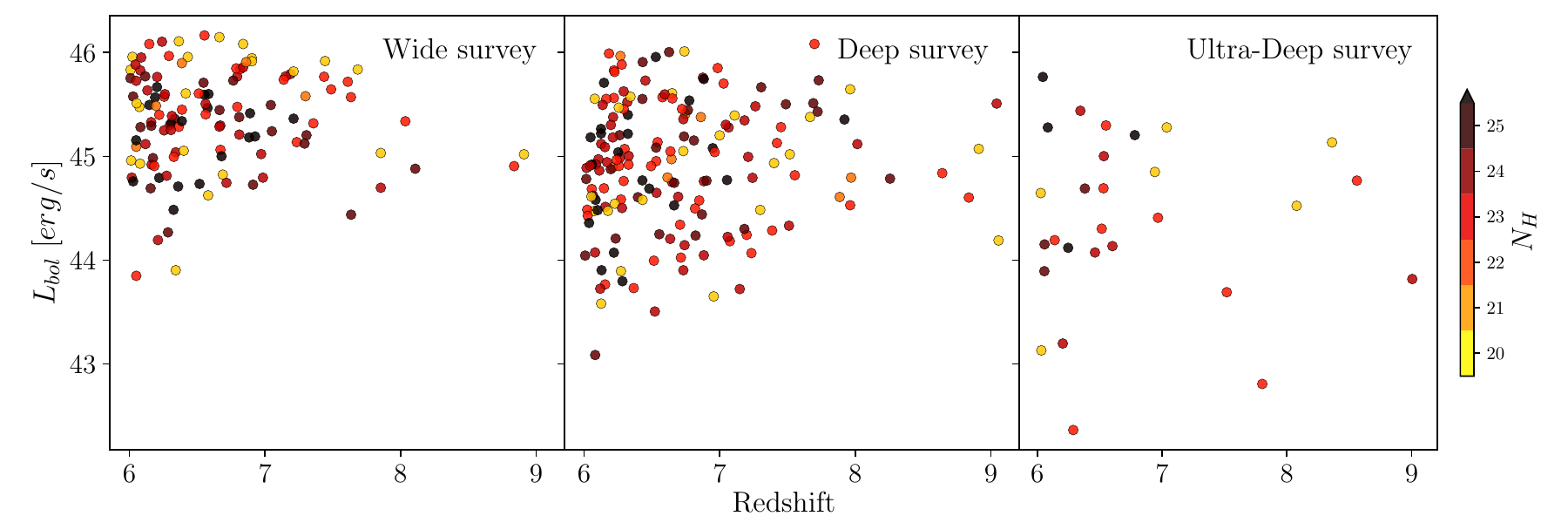}
    \caption{Simulated $z>6$ AGN to be detected by SKAO in the three tiers of its continuum 1.4GHz surveys. Each point is color-coded according to its obscuration level. The top three panels show the distribution of $\rm \log L_{1.4GHz}$  while the bottom panels show the distribution of $\rm \log L_{bol}$. The number of simulated sources in the Wide survey panels is almost 1/10 of the number of AGN reported in Tab.~\ref{tab:SKA_pred_v2} (see text for details).}
    \label{fig:SKA_pred}
\end{figure*}

By the end of this decade, the Square Kilometer Array Observatory (SKAO) will become fully operational. As reported in \cite{prandoni15}, SKAO will have among its primary science drivers the investigation of the SMBH-galaxy coevolution and the astrophysics related to the accretion processes. \cite{prandoni15} have proposed three 1.4GHz radio continuum surveys, optimized to address AGN/galaxy evolution science cases. These surveys are organized in tiers, following a nested wedding cake strategy; the Ultra-Deep, the Deep, and the Wide surveys with the following area -- $1\sigma$ sensitivity combinations: 0.05$\rm \mu Jy$ over 1 $\rm deg^2$, 0.2$\rm \mu Jy$ over 10-30 $\rm deg^2$, 1$\rm \mu Jy$ over $\rm 10^3 \ deg^2$, respectively. \\
Using the NC derived from our 1.4GHz model we computed the number of AGN and CTK AGN detectable in the three different survey tiers for high-redshift ($z>3$) and very high-redshift ($z>6$, $z>10$) ranges. The results for the three tiers are reported in Tab~\ref{tab:SKA_pred_v2}. All the predictions are computed assuming a flat sensitivity over the survey area. With the wide SKA survey thousands of AGN are expected to be detected at $z>6$ and a consistent fraction of them is expected to be CTK (at least $>45\%$). Also for $z>10$, a redshift range that since the advent of JWST  was almost completely unexplored, SKA will be able to detect several tens of AGN.\\
The  $z>6$ universe is still very unconstrained, particularly for the AGN X-ray and radio LFs. Without better information, we decided to use our baseline radio model and to extrapolate it to higher redshifts. However, we are aware that the results should be taken cautiously. For example, we assumed that the radio-to-X-ray relation measured by \citet{damato22} holds even at $z>>3$, but we did not consider whether the increased ISM density at high-z \citep{gilli22} may provide stronger free-free absorption and depress radio emission. Similarly, we did not consider whether the effects of self-synchro absorption (SSA) may depress radio emission as well, as one would require on average stronger magnetic fields to overcome energy losses by Comptonization of the enhanced CMB photon field. We defer a detailed discussion of these effects to future work entirely dedicated to SKA forecasts. Another uncertain aspect is the fraction of CTK AGN at these redshifts. While our $z>3$ baseline model assumes a constant fraction of CTK AGN, the large majority of the AGN above $z>6$ are expected to be CTK \citep{gilli22, ni20, lapi20, lambrides20, lusso23}.\\
At $z=6$ the $5\sigma$ sensitivities of the three survey correspond to 1.4GHz luminosities of $\rm 7.8\times 10^{38} erg\ s^{-1}$, $\rm 3.1\times 10^{39} erg\ s^{-1}$, $\rm 1.5\times 10^{40} erg\ s^{-1}$ respectively for the Ultra-Deep, Deep and Wide surveys.  
Converting the 1.4GHz luminosity into the corresponding HB X-ray intrinsic luminosity using Eq.~\ref{eq:LR_LX}, and taking the HX to bolometric luminosity correction factor of \cite{duras20}, we derived the minimum $z>6$ AGN bolometric luminosity that SKA would detect in each of the three surveys. These bolometric luminosity lower limits corresponds to $\rm L_{bol}\sim2\times 10^{44} erg\ s^{-1}$ for the Ultra-Deep, $\rm L_{bol}\sim10^{45} erg\ s^{-1}$ for the Deep and $\rm L_{bol}\sim8\times 10^{45} erg\ s^{-1}$ for the Wide survey. \\
To investigate the population of AGN that SKA will detect in the three surveys, we used the X-ray AGN mock catalog generated using the \cite{vito14} HXLF and provided in \cite{marchesi20}. This 100 deg$^2$ catalog includes $\sim 2.5\times 10^{6}$ AGN in the redshift range $3<z<20$ down to a 0.5-2keV luminsoity $\rm \log L_{SB}=40$, reaching fluxes below $\rm \sim 2\times 10^{-20} erg\ s^{-1}\ cm^{-2}$. Using the same relations as in Sec.~\ref{sec:radio_X?}, we derived the bolometric luminosity, the 1.4GHz luminosity and the flux density of each source.\\
The panels of Fig.~\ref{fig:SKA_pred} show the distribution in $\rm \log L_{1.4\rm GHz}$ and $\rm \log L_{bol}$ of the $z>6$ sources with a radio flux density larger than the sensitivity threshold of the three tiers.\\
Since the area of the mock catalog is 1/10 of the area of the Wide SKA survey, the number of sources in the Wide Survey panel is almost 1/10 of the AGN predicted for the respective survey at $z>6$ and reported in Tab.~\ref{tab:hz_pred}. 
The median bolometric luminosities of AGN detected in the three surveys, at $z>3$, z>6, and $z>10$ are reported in Tab.~\ref{tab:SKA_pred_v2}.\\ 
We remind that the $L_{1.4\rm GHz}-L_{HX}$ relation used to derive the radio luminosities from the mock X-ray catalog does not include the most powerful RL AGN that should populate a fainter region of the $\rm \log L_{bol}$ distribution. Furthermore, the upper limits of the distributions of the bolometric and radio luminosities presented in Fig \ref{fig:SKA_pred} are affected by the shape of the HXLF of \cite{vito14}. These HXLF are partially incomplete in their brightest part, since they were computed on deep pencil-beam X-ray fields that miss the most luminous X-ray sources. Taking into account a larger fraction of bright X-ray sources the $\rm \log L_{bol}$ distribution in Fig.~\ref{fig:SKA_pred} would extend to larger values. \\
\\
The possibility to identify thousands of AGN and CTK AGN at $z>6$ will transform our understanding of the co-evolution of MBHs with their galaxies in the first Gyr of the cosmic time.\\ 
The combination of radio data from SKA with multi-band data from current and next-generation facilities will be crucial to detect distant AGN and separate them from SFGs. Indeed SKA surveys are planned to overlap with forthcoming wide-area, sensitive optical and NIR-surveys, like those by LSST and Euclid (see \citealt{prandoni15} for a detailed list of the multi-wavelength survey synergies). Therefore, it will be possible to apply radio-excess (REX) selection techniques similar to those mentioned earlier in this work (see e.g. \citealt{Smoilcic17c} ) to very large statistical samples, and especially at high redshifts. In addition, in combination with ALMA follow-up data, SKA will prove whether the far-infrared-radio correlation normally used to separate AGN and SFGs is still valid at high-z, and also for low stellar masses and SFR. As an example, the sensitivity of the SKA deep surveys will allow detection of SFR $\rm \sim 10 M_{\odot}$/yr in galaxies up to $z\sim 3-4$, and SFR $\rm \geq50 M_{\odot}$/yr at $z\sim 6-7$ \citep{mcalpine15}. Observations of well-known extra-galactic fields where dense multi-band information, from UV to far-IR, is available will in particular increase the identification efficiency of CTK AGN. 
JWST and ELT observations will also allow for spectroscopic follow-up of radio-selected, high-z AGN candidates to determine their redshift and ultimately measure the radio AGN luminosity function at $z>6$. The combination of the SKA radio observations with future X-ray data from telescopes like STAR-X\footnote{http://star-x.xraydeep.org/}, Athena \citep{nandra13}, and AXIS \citep{mushotzky19} will enable improved AGN identification and measurement of ther column densities.\\
In the absence of other multi-band AGN diagnostics, the identification of AGN and CTK AGN using SKA data may rely on either multi-frequency radio information, or high-resolution (SKA 10GHz observations will provide angular resolution of $0.05"-0.1"$ compared to the $0.4"-0.5"$ of 1.4GHz observations) and/or multi-epoch follow-ups. All these diagnostics are amply used in the literature. Flat or convex radio spectral indices would point to the presence of compact AGN cores \citep{odea21}. High-resolution follow-ups can pinpoint AGN through the measurement of high ($T>10^{5-6}$ K) brightness temperatures \citep[see][]{morabito22}, while multi-epoch observations may identify the tiny fraction of variable AGN at $\mu$Jy flux density levels \citep{radcliffe19b}. 

\section{Conclusions}
In this work we developed an analytical model to compute the 1.4GHz luminosity function and associated number counts for the whole AGN population and for subpopulations of AGN with different obscuration levels. In particular, we converted into the radio band the prescriptions of the population synthesis model of the CXB using an X-ray to radio luminosity relation derived for faint X-ray sources.\\
We applied our model to some of the major extragalactic fields covered by deep radio and X-ray observations to predict the number of AGN and CTK AGN detectable in each band.\\
The main results can be summarized as follows:
   \begin{enumerate}  
      \item We found a very good agreement between the number of radio-detected AGN predicted by our model in the fields and the observed number of AGN identified via multiple selection techniques. This means that our model is able to give an almost complete census of the radio AGN population, excluding the minority population of RL AGN, that is missed by construction.
      \item On average X-rays are able to detect a larger number of AGN, but most of the X-ray detected AGN are unobscured. The CTK AGN surface density detected at 1.4GHz is on average $10$ times higher than the X-ray one. \\
      This result stands also at high redshift ($z>3$), where the surface density of CTK AGN expected in radio surveys becomes, for some fields, even $10^3$ times larger than the X-ray density.
      \item Our model predicts the existence of thousands of CTK AGN already detected in the radio fields investigated in this work that are largely missed by the corresponding X-ray observations. Both our model and the results coming from the literature suggest that radio emission may provide an unbiased picture (in terms of obscuration) of the AGN demography. However, multiple selection techniques employing different multi-wavelength indicators of nuclear activity are required to identify AGN among SFGs. The investigation of the CTK AGN selection among the sources in the available radio catalogs will be the focus of the next-coming work.
      \item In the future the SKAO will be able to detect in its three-tier surveys show that it will detect more than 2000 AGN at $z>6$ (down to $\rm \log L_{bol}\sim 43$) and some tens at $z>10$, opening new windows in exploration of the AGN parameter space at these redshifts.     
\end{enumerate}
In this work we demonstrated that radio emission can be a powerfull tool to detect the elusive population of heavily obscured, Compton-thick AGN. In the future, the synergies between deep continuum surveys performed by SKA and multi-band information provided by present and future telescopes and observatiories, will allow for an extensive testing of radio-based AGN selection tecniques on large statistical samples, and for a detailed exploration of the radio emission properties of these objects in the high-z Universe.

\begin{acknowledgements}
We thank the anonymous referee for useful suggestions which improved the quality of the paper.\\
We acknowledge financial support from the grant PRIN MIUR 2017PH3WAT (‘Black hole winds and the baryon life cycle of galaxies’). IP and RG acknowledge support from INAF under the Large Grant 2022 funding scheme (project "MeerKAT and LOFAR Team up: a Unique Radio Window on Galaxy/AGN co-Evolution")".
We acknowledge useful discussion with I. Delvecchio, T. Costa, F. La Franca. 
      
\end{acknowledgements}

%
%

\bibliographystyle{aa}
\bibliography{p1}

\end{document}